\documentclass[12pt]{article}

\usepackage{graphics}

\usepackage{cite} 
\usepackage{epsfig}
\newenvironment{Eqnarray}%
     {\arraycolsep 0.14em\begin{eqnarray}}{\end{eqnarray}}

\makeatletter
\@addtoreset{equation}{section}
\def\theequation{\thesection.\arabic{equation}}
\makeatother

\begin{document}
\def\nicefrac#1#2{\hbox{${#1\over #2}$}}

\setlength{\textwidth}{16.8 cm}
\setlength{\textheight}{23 cm}
\addtolength{\topmargin}{-2.5 cm}

\newcommand{\nn}{\nonumber}
\newcommand{\raw}{\rightarrow}
\newcommand{\be}{\begin{equation}}
\newcommand{\ee}{\end{equation}}
\newcommand{\bea}{\begin{Eqnarray}}
\newcommand{\eea}{\end{Eqnarray}}
\newcommand{\dl}{\stackrel{\leftarrow}{D}}
\newcommand{\dr}{\stackrel{\rightarrow}{D}}
\newcommand{\dd}{\displaystyle}
\newcommand{\slas}[1]{\rlap/ #1}
\def\lsim{\mathrel{\raise.3ex\hbox{$<$\kern-.75em\lower1ex\hbox{$\sim$}}}}
\def\gsim{\mathrel{\raise.3ex\hbox{$>$\kern-.75em\lower1ex\hbox{$\sim$}}}}
\def\ifmath#1{\relax\ifmmode #1\else $#1$\fi}
\def\beq{\begin{equation}}
\def\eeq{\end{equation}}
\def\beaa{\begin{array}}
\def\eeaa{\end{array}}
\def\ts{\tilde{t}}
\def\cs{\tilde{c}}

\def\ls#1{\ifmath{_{\lower1.5pt\hbox{$\scriptstyle #1$}}}}

%%%%%%%%%%%%%%%%%%%%%%%%%%%%%%%%%%%%%%%%%%%%%%%%%%%%%%%%%
% Preprint cover page material
%%%%%%%%%%%%%%%%%%%%%%%%%%%%%%%%%%%%%%%%%%%%%%%%%%%%%%%%%
\pagestyle{empty}
\begin{flushright}
FTUAM/02-21 \\
IFT/CSIC-UAM 02-37 \\
\end{flushright}
\vskip1cm

\renewcommand{\thefootnote}{\fnsymbol{footnote}}
\begin{center}
{\Large\bf
Flavour Changing Neutral Higgs Boson Decays from Squark - Gluino Loops}\\[1cm]
{\large 
Ana M. Curiel,
Mar\'{\i}a J. Herrero 
and David Temes
~\footnote{electronic addresses:
curiel@delta.ft.uam.es, herrero@delta.ft.uam.es,
temes@delta.ft.uam.es}
}\\[6pt]
{\it Departamento de F\'{\i}sica Te\'{o}rica \\
   Universidad Aut\'{o}noma de Madrid,
   Cantoblanco, 28049 Madrid, Spain 
.}
\\[1cm]

\begin{center}
{\bf Abstract}
\end{center}
\end{center}

 We study the flavour changing neutral Higgs boson decays that can be induced
 from genuine supersymmetric particles at the one-loop level and 
 within the context
 of the Minimal Supersymmetric Standard Model. We consider all the possible 
 flavour changing decay channels of the three neutral Higgs bosons 
 into second and third generation quarks, and
 focus on the Supersymmetric-QCD corrections from squark-gluino loops which are
 expected to provide the dominant contributions. We assume here the more general
 hypothesis for flavour mixing, where there is 
 misalignment between the quark and squark sectors, leading to a 
 flavour non-diagonal squark mass matrix.  The form factors involved, and
 the corresponding Higgs partial decay widths and branching ratios, are computed
 both analytically and numerically, and their behaviour with the parameters 
 of the Minimal Supersymmetric Standard Model and with the squark mass mixing 
 are analyzed in full detail. The large rates found, are explained in terms of
 the non-decoupling behaviour of these squark-gluino loop corrections in
 the scenario with very large supersymmetric mass parameters. Our results
 show that if these decays are seen in future colliders they could provide 
 clear indirect signals of supersymmetry.

\vfill
\clearpage
%%%%%%%%%%%%%%%%%%%%%%%%%%%%%%%%%%%%%%%%%%%%%%%%%%%%%%%%%%%%%%%%%%%%%%%%

\renewcommand{\thefootnote}{\arabic{footnote}}
\setcounter{footnote}{0}

%%%%%%%%%%%%%%%%%%%%%%%%%%%%%%%%%%%%%%%%%%%%%%%%%%%%%%%%%%%%%%%%%%%%%%%%%%%%
\pagestyle{plain}
\section{Introduction}
\label{chap.introduction}
 The indirect searches of supersymmetric (SUSY) particles via their radiative 
 effects into low energy observables have received a lot of attention in the 
 last decades~\cite{pbmz-polonsky,proceedings}.
  The main motivation for these searches is that
 they can provide valuable clues in the way towards the final discovery 
 of supersymmetry, even in the most pessimistic scenario where the SUSY 
 spectrum
 is too heavy as to be directly produced in the present or forthcoming
 colliders. In this concern, the flavour-changing neutral current (FCNC)
 processes are 'ideal laboratories' where to look for these indirect SUSY
 signals or any other radiative effects from new physics beyond the Standard
 Model (SM) of particle interactions, since the SM predicts negligible rates 
 for these processes. We are interested here in the FCNC
 effects that are generated from genuine SUSY radiative corrections 
 within the context of the Minimal Supersymmetric Standard Model (MSSM)
 \cite{Haber:1985rc}.
 Our interest is focused particularly on the SUSY radiative effects 
 that induce scalar flavour changing interactions~\cite{9604378}.
    
 The strong suppression of any FCNC process within the SM is due to the absence
 of tree-level flavour changing (FC) interactions and the GIM cancellation mechanism
\cite{GIM} that operates beyond the tree-level. Within the MSSM, the scalar 
 FC interactions are also absent at tree-level, but they can be generated 
quite efficiently at the one-loop level and lead to sizeable contributions at 
some
regions of the MSSM parameter space. This is mainly because the GIM suppression
mechanism  does not necessarily operate in the genuine SUSY radiative
corrections~\cite{Gatto}.  Special mention 
deserves the SUSY 
one-loop radiative effects that modify the tree-level relations between the 
down-type
fermion mass matrices and their corresponding Yukawa coupling matrices~\cite{inter}, 
since they can induce FC Yukawa interactions if there is change of 
flavour in
the internal squark lines~\cite{toharia,kolda}. This has been the 
subject of numerous studies in the last years, because these SUSY radiative
effects are very
significant at large $\tan\beta$ values \cite{inter,toharia,kolda}. 
This $\tan\beta$ reassures the ratio 
of the vacuum expectation values of the two Higgs doublets and is predicted 
to be large in some SUSY-GUT models with top-bottom Yukawa coupling 
unification,
being of order $m_t/m_b\sim 50$~\cite{largetb}. On the other hand, the phenomenological interest
of these $\tan\beta$ enhanced SUSY
radiative corrections leading to FC neutral Higgs 
interactions is that they affect a wide range of low energy mesonic 
processes and can be in conflict with present experimental data,
including $B^0-\bar B^0$ mixing \cite{toharia,chanko1,isidori,newchanko}, leptonic $B$ meson decays
\cite{kolda,chinos,oldchanko,alemanes1,alemanes2,newisidori}, as well as some other CP-conserving
\cite{ellis,logan} and CP-violating observables related to 
 the $K$ and $B$ meson physics \cite{dedes}. Similar
enhanced $\tan\beta$ effects~\cite{bstanbeta} 
have 
been found
at $b \rightarrow s \gamma$ decays~\cite{bsgamma} as well.

Another studies at higher energy processes of FC scalar 
interactions being induced from 
SUSY loops 
have also been performed in the literature. These include rare $Z$ boson decays 
\cite{duncanZ,Zbs} and rare top decays~\cite{sola,mele,He,nueva}, both 
being significant at particular regions of the MSSM parameter space.            
Specially large are the branching ratios for the 
$Z \rightarrow b \bar s$ decays, which can reach $10^{-6}$ for 
large $\tan\beta$ 
values~\cite{Zbs} and  for the top rare decay into neutral MSSM Higgs bosons, 
$t \rightarrow c H$, which can be as large as $10^{-4}$~\cite{sola,He}, and both
 rates are
largely dominated by the SUSY radiative effects from squark-gluino loops. 
 With such a large rates, these $Z\rightarrow b \bar s$ decays could be accessible at a future Giga-Z 
 collider~\cite{Zbs},
 and the $t \rightarrow c H$ decays could be observed at the 
 forthcoming CERN-LHC
 collider~\cite{aguilar}. 
    
 Our subject of study here is the flavour changing neutral Higgs boson decays 
 (FCHD) that can be induced from genuine SUSY particles at the one-loop level.
 This is a subject closely related to the previous ones but, to our knowledge,
 has not been analyzed in the literature yet. We will work here in the MSSM
 context and consider all the possible FC decay channels of the
 three neutral MSSM Higgs bosons into second and third generation quarks,
 namely, $H  \rightarrow b\bar s,s\bar b,t\bar c,c \bar t$, with 
 $H =h_o,H_o,A_o$ and, as usual, $h_o$ and $H_o$ are the lightest and
 heaviest CP-even bosons, respectively, and $A_o$ is the CP-odd boson. 
 Our study is devoted to the second and third generation quarks because the
 squark mixing between these two generations, which is been assumed here to 
 induce these
 decays, is the less constrained experimentally~\cite{9604378}. We have 
 focused 
 on the SUSY radiative corrections from the SUSY-QCD sector which, by analogy 
 to the previous studies, are expected to provide the dominant contributions 
 to the FCHD rates, as their strength is driven by the strong coupling constant.
 
 More concretely, we will compute here the one-loop contributions of 
 ${\cal O}(\alpha_S^2)$ to the partial
 decay widths of the neutral Higgs bosons coming from loops of gluinos and 
 third 
 and second generation squarks. We will present the analytical
 results from our diagrammatic computation in terms of the involved form
 factors and we will analyze in full detail the numerical values of the
 Higgs partial widths as a function of various relevant MSSM parameters. Basically, 
 the $\mu$ mass parameter, the gluino mass, the squark masses, the pseudoscalar 
 mass $m_{A}$ and $\tan \beta$.
 
 Regarding our hypothesis on the generation of squark flavour mixing, 
 we consider here the most general scenario, which is called in the literature
 non-minimal flavour scenario. This occurs when the squark mass matrices are 
 not flavour-diagonal in the same basis as the quark mass matrices. Thus, 
 when the
 squark mass matrices are diagonalized, FC gluino-quark-squark
 couplings arise for the mass eigenstates, and these induce in turn the FCHD 
 via squark-gluino loops, which are the subject of our study. It is worth to
 recall that this hypothesis of misalignment between the squark and quark
 mass matrices is present in the most general parameterization of the MSSM and
 can lead to dangerous FC effects in conflict with experiment. 
 Specially, the data on $K^0-\bar K^0$ and $D^0-\bar D^0$ mixing impose severe constraints
 on the mixing involving the first generation~\cite{9604378}. This is why we focus on the
 mixing between the second and third squark generations, which is practically 
 unconstrained~\cite{9604378}.
 
 Attempts to solve the previously commented SUSY flavour problem of the MSSM include flavour-blind
 SUSY breaking scenarios, as in Minimal Supergravity, in which the sfermion mass
 matrices are flavour diagonal in the same basis as the quark mass matrices at
 the SUSY-breaking scale, but a small amount of non-minimal flavour mixing is
 generated as the squark and quark masses are evolved down, 
 via renormalization group equations, to the electroweak scale. The actual
 estimates of these radiatively induced flavour off-diagonal squark squared mass
 terms~\cite{Hikasa,RGE_supergravity} indicate that the largest ones are those referred to the SUSY 
 partners of
 the left-handed quarks, $\Delta_{LL}$, since these scale with the squared of 
 the soft
 SUSY breaking masses, in contrast to the $\Delta_{LR}$ 
 (or $\Delta_{RL}$) and $\Delta_{RR}$ terms
 that scale with one or zero powers respectively of the soft SUSY breaking
 masses. Thus, the hierarchy $\Delta_{LL}>>\Delta_{LR}>>\Delta_{RR}$ is usually
 assumed. These same estimates also indicate that the 
  $\Delta_{LL}$ terms generating the mixing between the second
  and third generation squarks can be numerically significant because of the 
  third generation quark mass factors involved. 
  
  In our analysis of the FCHD we will assume the simplest and 
  theoretically better motivated hypothesis, where the only non-zero
  off-diagonal squark squared mass entries in the $\tilde d$-sector and 
  $\tilde u$-sector 
  are for $\tilde {s_L} \tilde {b_L}$ and 
  $\tilde {c_L} \tilde {t_L}$ mixing, respectively. We will parameterize these 
  flavour off-diagonal entries, {\it a la} Sher~\cite{AnsatzSher}, simply by 
  $\Delta_{LL}=\lambda_{LL}M_{\tilde Q}M_{\tilde Q'}$, where  
   $M_{\tilde Q}$ and $M_{\tilde Q'}$ are the two corresponding involved 
   soft SUSY breaking masses. The dimensionless parameter 
   $\lambda_{LL}$ is considered here to be the only free-parameter
   characterizing the flavour mixing strength and, for the numerical 
   evaluations, it
   will be taken in the conservative range, $0\le \lambda_{LL} \le 1$, which is
   perfectly allowed by the present data (for a summary of present 
   bounds on the 
   $\lambda_{LR}$, $\lambda_{RL}$, $\lambda_{LL}$ and $\lambda_{RR}$
   parameters, see~\cite{9604378}).

  The paper is organized as follows. We present in 
  section~\ref{FC_squarksector}  the flavour non-diagonal squark   
squared mass matrices and write the relevant 
quark-squark-gluino interaction terms that are generated after rotating to the
mass eigenstate basis. The computation of the one-loop corrections to 
the
form factors and FCHD widths and their analytical results are presented 
in section~\ref{chap.SUSY-QCD_cont}.
The numerical analysis of the FCHD rates and a detailed discussion on their 
dependence with the MSSM parameters and with the $\lambda_{LL}$ parameter 
are included in 
section~\ref{chap.phenomenology}. After scanning the MSSM parameter space,
particular regions where the FCHD rates are maximal are detected. The
values of the FCHD branching ratios as a function of 
the $\lambda_{LL}$ parameter are then analyzed at these regions.
 We devote section~\ref{chap.decoupling} to the study of the
SUSY decoupling properties in these FC observables. After
performing a large SUSY mass expansion,
 we will show in this section 
 that the SUSY
radiative effects from squark-gluino loops indeed do not decouple in the FCHD.
This explains why the numerical size of the FCHD rates found in this work are 
so large. Finally, the last section is devoted to the summary and conclusions.

%%%%%%%%%%%%%%%%%%%%%%%%%%%%%%%%%%%%%%%%%%%%%%%%%%%%%%%%%%%%%%%%%%%%%%%%%
\section{Flavour Changing in the SUSY-QCD sector of the MSSM}
\label{FC_squarksector} 

As recalled in the introduction, in the MSSM there is a new source of flavour changing phenomena coming from a possible 
misalignment between the rotation that diagonalizes the quark and squark sectors. The squark mass matrix,
expressed in the basis in which the squarks fields are rotated parallel to the
quarks (called the super-CKM basis), is in general
non-diagonal in flavour space, and provides new FC effects. 

Assuming that FC 
squark mixing is significant only in transitions between the third and second
generation
squarks, and that there is only LL mixing, given by the Sher type 
ansatz~\cite{AnsatzSher}, the squark squared mass matrices 
in the ($\tilde c_L$,$\tilde c_R$,$\tilde t_L$,$\tilde t_R$) and 
($\tilde s_L$,$\tilde s_R$,$\tilde b_L$,$\tilde b_R$)
 basis, respectively, can be written as follows,

\beq
M^2_{\tilde u} =\left\lgroup 
         \beaa{llll}
          M_{L,c}^2  &  m_c X_c & \lambda_{LL} M_{L,c} M_{L,t}& 0\\
          m_c X_c   &  M_{R,c}^2  & 0  &  0\\
          \lambda_{LL} M_{L,c} M_{L,t} & 0 & M_{L,t}^2  &  m_t X_t\\
          0 & 0 & m_t X_t &  M_{R,t}^2 
\eeaa
         \right\rgroup 
\label{eq.usquarkmass}
\eeq
\beq     
M^2_{\tilde d} =\left\lgroup 
         \beaa{llll}
          M_{L,s}^2  &  m_s X_s & \lambda_{LL} M_{L,s} M_{L,b}& 0\\
          m_s X_s   &  M_{R,s}^2  & 0  &  0\\
          \lambda_{LL} M_{L,s} M_{L,b} & 0 & M_{L,b}^2  &  m_b X_b\\
          0 & 0 & m_b X_b &  M_{R,b}^2
\eeaa
         \right\rgroup
\label{eq.dsquarkmass}
\eeq
where
\begin{eqnarray}
M_{L,q}^2 &=& M_{\tilde Q,q}^2 +m_q^2 + \cos2\beta (T_3^{q}-Q_q s_W^2)m_Z^2\, , \nn \\
M_{R,(c,t)}^2 &=& M_{\tilde U,(c,t)}^2 +m_{c,t}^2 + \cos2\beta Q_t s_W^2 m_Z^2\, , \nn \\
M_{R,(s,b)}^2 &=& M_{\tilde D,(s,b)}^2 +m_{s,b}^2 + \cos2\beta Q_b s_W^2 m_Z^2\, , \nn \\
X_{c,t} &=& m_{c,t} ( A_{c,t} - \mu \cot \beta) \, , \nn \\
X_{s,b} &=& m_{s,b} ( A_{s,b} - \mu \tan \beta) \,; \nn \\
\label{eq.MU4x4_param}
\end{eqnarray}
$m_q$, $T_3^q$, $Q_q$ are the mass, isospin and electric charge respectively of the quark 
$q$;  $m_Z$ is the $Z$ boson mass, and $s_W$ is the sine of the weak angle
$\theta_W$. 
The involved MSSM parameters in the SUSY-QCD sector are, 
as usual, the gluino mass, $M_{\tilde g}$, the $\mu$-parameter, the soft
SUSY breaking masses $M_{\tilde Q}$, $M_{\tilde U}$, $M_{\tilde D}$ and the
soft SUSY breaking trilinear parameter, $A$. Notice that we have used 
in eqs.(\ref{eq.usquarkmass}), (\ref{eq.dsquarkmass}) and 
(\ref{eq.MU4x4_param}) a self-explanatory notation for the flavour indices, and  
that, due to the $SU(2)_L$ invariance, 
$M_{\tilde Q,c} = M_{\tilde Q,s}$ and $M_{\tilde Q,t} = M_{\tilde Q,b}$.

 In our previous parameterization of flavour mixing in the squark sector, there
 is only one free-parameter, the $\lambda_{LL}$ parameter, that characterizes the
 flavour mixing strength.  Notice that, for the sake of simplicity, 
 we are assuming the
 same notation for the $\lambda_{LL}$ parameter 
 in the $\tilde t-\tilde c$  and  $\tilde b - \tilde s$ sectors,
 which
 from now on will be called  
 simply $\lambda$. For the numerical estimates in the following sections,
 the $\lambda$ values   
  will be taken in the moderate range $0\le \lambda \le 1$. Obviously, 
  the choice
 $\lambda=0$ represents the zero-flavour mixing case.   
   
In order to diagonalize the two previous 
$4 \times 4$ squark mass matrices, two rotation $4 \times 4$ matrices, 
$R^{(u)}$ and
$R^{(d)}$, one for the $up$-type squarks and the other one for the $down$-type 
squarks, are needed. Thus, the squark mass eigenstates, 
$\tilde q_{\alpha}$, 
and the interaction squark eigenstates, $\tilde q'_{\alpha}$, are related by, 
\begin{eqnarray}
\tilde q'_{\alpha} &=& \sum R_{\alpha \beta}^{(q)} \tilde q_{\beta} \, ,
\end{eqnarray}
where now the notation is,
\begin{eqnarray}
\tilde u'_{\alpha} 
= \left( \begin{array}{c}  \tilde c_L \\ \tilde c_R \\ \tilde t_L \\ \tilde t_R  \end{array}  \right)
\, \, , \, \,
\tilde d'_{\alpha} 
= \left( \begin{array}{c}  \tilde s_L \\ \tilde s_R \\ \tilde b_L \\ \tilde b_R  \end{array}  \right)
&,&
\tilde u_{\beta} 
= \left( \begin{array}{c}  \tilde u_1 \\ \tilde u_2 \\ \tilde u_3 \\ \tilde u_4  \end{array}  \right)
\, \, , \,\,
\tilde d_{\beta} 
= \left( \begin{array}{c}  \tilde d_1 \\ \tilde d_2 \\ \tilde d_3 \\ \tilde d_4  \end{array}  \right)
\, 
\end{eqnarray}

Once the squark mass matrices are diagonalized, one obtains the squark mass 
eigenvalues
and eigenstates that obviously depend on $\lambda$. Again, the 
choice $\lambda=0$ 
represents the case where no flavour mixing occurs, and recovers the usual
pairs of physical 
squarks, ($\tilde b_1, \tilde b_2$),  
($\tilde s_1, \tilde s_2$), ($\tilde t_1, \tilde t_2$), 
and ($\tilde c_1,\tilde c_2$) with the usual mass patterns. 

For illustrative purposes, we show in fig.\ref{upsqmass_lambda} the physical
masses of the $down$-type squarks, $M_{\tilde d_{1,2,3,4}}$, as a function of $\lambda$, where, 
for definiteness, we have chosen all the SUSY mass parameters equal 
to 1 $TeV$ and $\tan \beta = 40$.  We can see clearly the numerical 
differences among the 
four squark masses 
$M_{\tilde q_{1,2,3,4}}$ (we take the convention 
$M_{\tilde q_1}>M_{\tilde q_2}>M_{\tilde q_3}>M_{\tilde q_4}$) and the  
corresponding ones for 
$\lambda=0$ that do not include any flavour changing. 
 In the $\tilde d$ sector, there are two eigenvalues, 
$M_{\tilde d_{2}}$ and $M_{\tilde d_{3}}$ that do not change much as we 
increase the value of $\lambda$. These correspond, when $\lambda =0$, to 
$\tilde s_1$ and 
$\tilde s_2$ respectively. On the other hand, $M_{\tilde d_{1}}$ increases 
with $\lambda$ while $M_{\tilde d_{4}}$ decreases. These last ones correspond, 
when $\lambda=0$, to $\tilde b_1$ and $\tilde b_2$ respectively. 

\begin{figure}[h]
\begin{center}
\epsfig{file=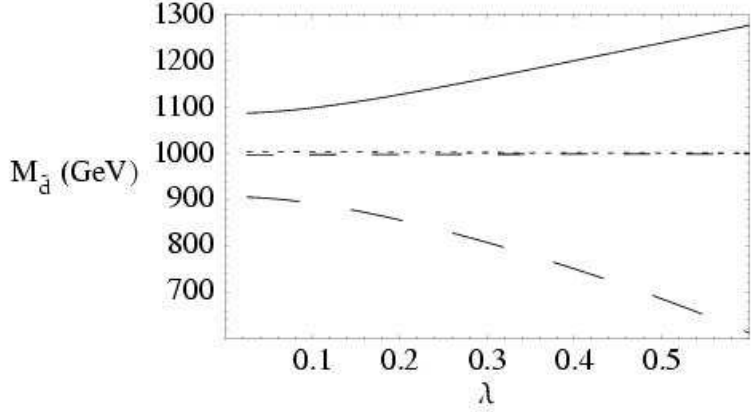,height=2.5in}
\caption{Physical masses of the $down$-type squarks as a function 
of $\lambda$ when all the SUSY mass parameters are equal to 1~$TeV$
and $\tan \beta = 40$. The solid line corresponds to $M_{\tilde d _1}$ , 
the long dashed line corresponds to $M_{\tilde d _4}$ and the other two 
correspond to $M_{\tilde d _2}$ and $M_{\tilde d _3}$ 
(short-dashed and semi-short-dashed) respectively.}
\label{upsqmass_lambda}
\end{center}
\end{figure}

The numerical analysis for the {\it up}-type squarks is completely 
analogous to the previous one of the {\it down}-type squarks and we
dont show it here explicitly. Concerning the mass pattern and its
dependence with $\lambda$, there are two eigenvalues, 
$M_{\tilde u_{2}}$ and $M_{\tilde u_{3}}$ that do not change much 
as we increase the value of $\lambda$, and that correspond, when 
$\lambda =0$, to $\tilde c_1$ 
and $\tilde c_2$ respectively. On the other hand, $M_{\tilde u_{1}}$ increases 
with $\lambda$ while $M_{\tilde u_{4}}$ decreases and these last ones 
correspond, when $\lambda=0$, to $\tilde t_1$ and $\tilde t_2$ respectively.

Finally, notice that, for the numerical analysis of the FCHD rates in the next
sections, only   $\lambda$ values in the 
$0 \le \lambda \le 1$ range that lead to physical squark masses above $150 \, GeV$
will be considered. The present experimental lower mass bounds on the 
squark masses of the first and second  squark generation
are actually even more stringent that this value~\cite{pdg2002}, 
but we have chosen here this common value of $150 \, GeV$ for 
simplicity and definiteness. Similarly, in view of the present experimental
lower bounds on the gluino mass~\cite{pdg2002}, we will consider here $M_{\tilde g}$ 
values above $200 \, GeV$.

The previously introduced intergenerational mixing effects in the squark sector
 lead to strong FC effects in processes involving 
neutral currents through the quark-squark-gluino interaction Lagrangian, 
which can now be written in the squark mass eigenstate basis
as,
\begin{eqnarray}
\mathcal{L} (\tilde g, \tilde q, q) &=& - \sqrt{2} g_s \mathtt{t}
\left( R_{3\alpha}^{(u)*} \bar{\tilde g} \tilde u_{\alpha}^* t_L + 
R_{3\alpha}^{(d)*} \bar{\tilde g} \tilde d_{\alpha}^* b_L
-  R_{4\alpha}^{(u)*} \bar{\tilde g} \tilde u_{\alpha}^* t_R 
- R_{4\alpha}^{(d)*} \bar{\tilde g} \tilde d_{\alpha}^* b_R \right. \nn \\
&& \left. + R_{1\alpha}^{(u)*} \bar{\tilde g} \tilde u_{\alpha}^* c_L + 
R_{1\alpha}^{(d)*} \bar{\tilde g} \tilde d_{\alpha}^* s_L
-  R_{2\alpha}^{(u)*} \bar{\tilde g} \tilde u_{\alpha}^* c_R 
- R_{2\alpha}^{(d)*} \bar{\tilde g} \tilde d_{\alpha}^* s_R 
\right) + h.c \nn \\
\label{eq.gluinosqq}
\end{eqnarray}
with $\alpha =1,2,3,4$ and we have used the short notation $\mathtt{t} \bar {\tilde g} \tilde q^* q \equiv t_{ij}^a \bar
{\tilde g}_a \tilde q^{*i} q^j$, where 
$t_{ij}^a$ are the standard $SU(3)_c$ generators. For simplicity, we will omit the color
indices from now on.  

In this way, if there is misalignment between quark and squarks, 
new FC effects will appear
in neutral currents processes and, in particular, 
in the neutral Higgs decays of the MSSM into quarks that we are interested 
in. These will occur, mainly, via loops of squarks and gluinos  and through 
the flavour non-diagonal $q-\tilde q-\tilde g$ couplings of 
eq.(\ref{eq.gluinosqq}) which in turn have emerged from the 
flavour non-diagonal squark squared mass matrices of eqs.(\ref{eq.usquarkmass})
and (\ref{eq.dsquarkmass}).
 These effects will be 
driven by the strong coupling constant $\alpha_S$, and therefore 
they are expected to be numerically large. We will study those effects in full detail in the 
forthcoming sections~\ref{chap.SUSY-QCD_cont},~\ref{chap.phenomenology} 
and~\ref{chap.decoupling}.

%%%%%%%%%%%%%%%%%%%%%%%%%%%%%%%%%%%%%%%%%%%%%%%%%%%%%%%%%%%%%%%%%%%%%%%%%%%%%
\section{Generating flavour changing Higgs decays from squark-gluino loops}
\label{chap.SUSY-QCD_cont}

We present in this section the computation of 
the one-loop radiative corrections to the 
FCHD of the neutral MSSM Higgs bosons 
into second and third generation quarks, 
$H \to b \bar s, s \bar b, t \bar c, c \bar t$. Notice that, the present 
theoretical upper mass bound 
$m{_{h_o}}\le 135 \, GeV$ \cite{loopmass} makes the lightest Higgs boson $h_o$ unable to decay into 
$t\bar c$ or $c \bar t$. Therefore, just the following channels will be
considered, $H_a \to b \bar s, s \bar b$ with $H_a=h_o,H_o,A_o$ and 
$H_a \to t \bar c, c \bar t$  with $H_a=H_o,A_o$.
We will focus on the radiative corrections that come from 
the SUSY-QCD sector, and more specifically those from loops of squarks 
of type $\tilde s$, $\tilde c$, $\tilde b$ and  $\tilde t$, and 
gluinos $\tilde g$. These will provide contributions to the FC
partial widths of $\mathcal{O} (\alpha_S^2)$ and are expected to be the 
dominant ones.

Notice that, 
the computation of these FC partial widths  
are relatively  easy and do not require
renormalization, because these decay processes do not proceed
at the tree level in the MSSM. One just computes the different
 one-loop diagrams
that contribute to the process and the final result obtained, 
after adding up all of them, must be
finite, since no lowest order interaction could absorb the left over infinities. 

In order to present our results for the partial decay widths, 
it is convenient to define the one-loop effective interaction term associated to each decay $H \to q \bar q'$ in the following compact form:

\begin{equation}
i F = -ig \bar u_q (p_1) (F_L^{qq'} (H) P_L + F_R^{qq'} (H) P_R) v_{q'}(p_2) H(p_3),
\end{equation}
where $g$ is the $SU(2)_L$ gauge coupling and $F_L$ and $F_R$ are the 
form factors associated to each chirality projection $L$ and $R$ respectively.
 The values of $F_L$ and $F_R$ follow from the explicit calculation of the one-loop vertices and 
 mixed self-energies and are of order $\alpha_S$. These one-loop diagrams from the SUSY-QCD sector 
are depicted in fig.\ref{fig.diagrams1} for the $H \to b \bar s$ decays, 
with $H=h_o,H_o,A_o$. The corresponding diagramas for the $tc$ decays
are identical but replacing $d \to u$, $b \to t$ and $s \to c$. 
   
\begin{figure}[h]
\epsfig{file=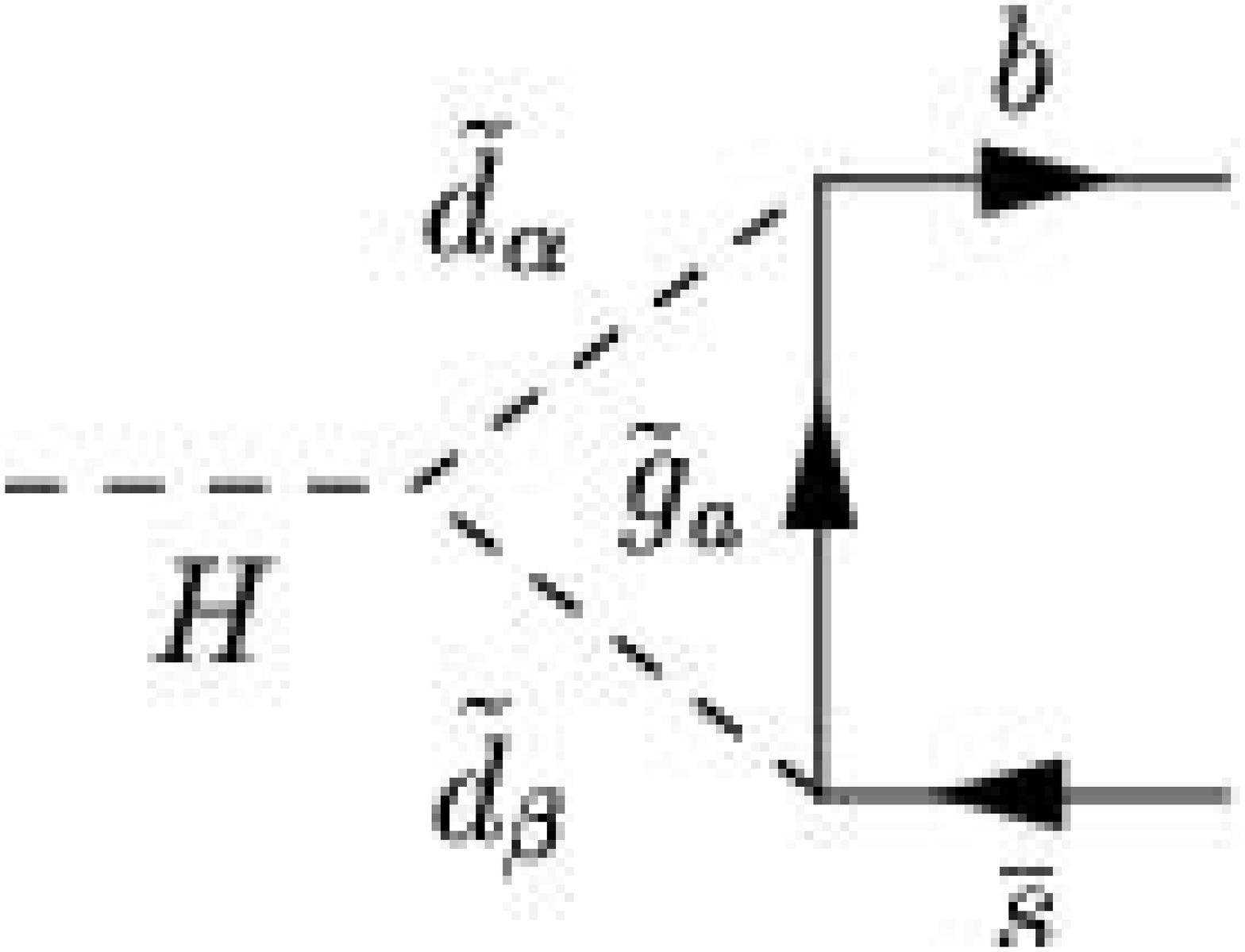,height=1.2in}
\hspace{0.025in}
\epsfig{file=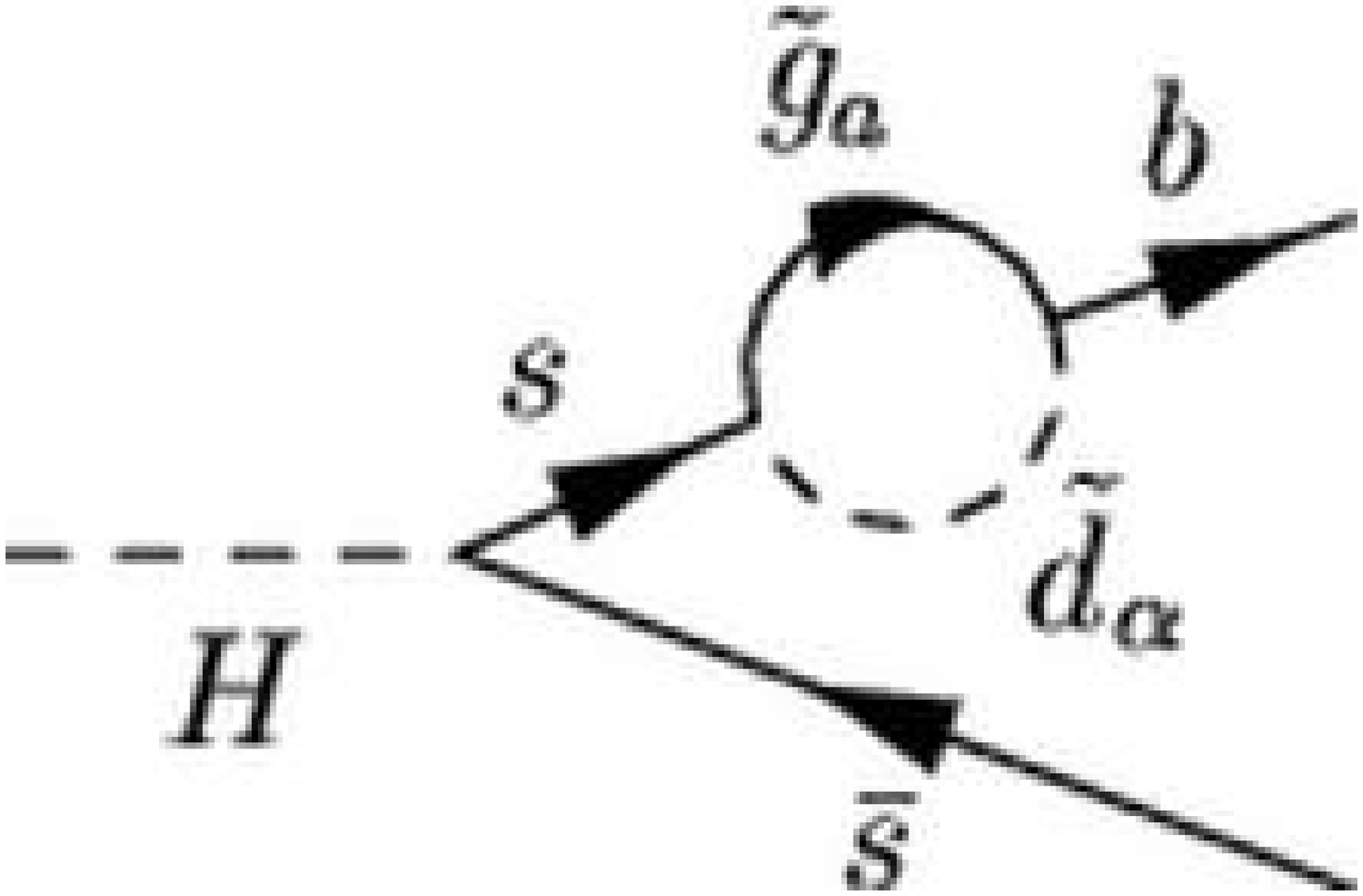,height=1.2in}
\hspace{0.025in}
\epsfig{file=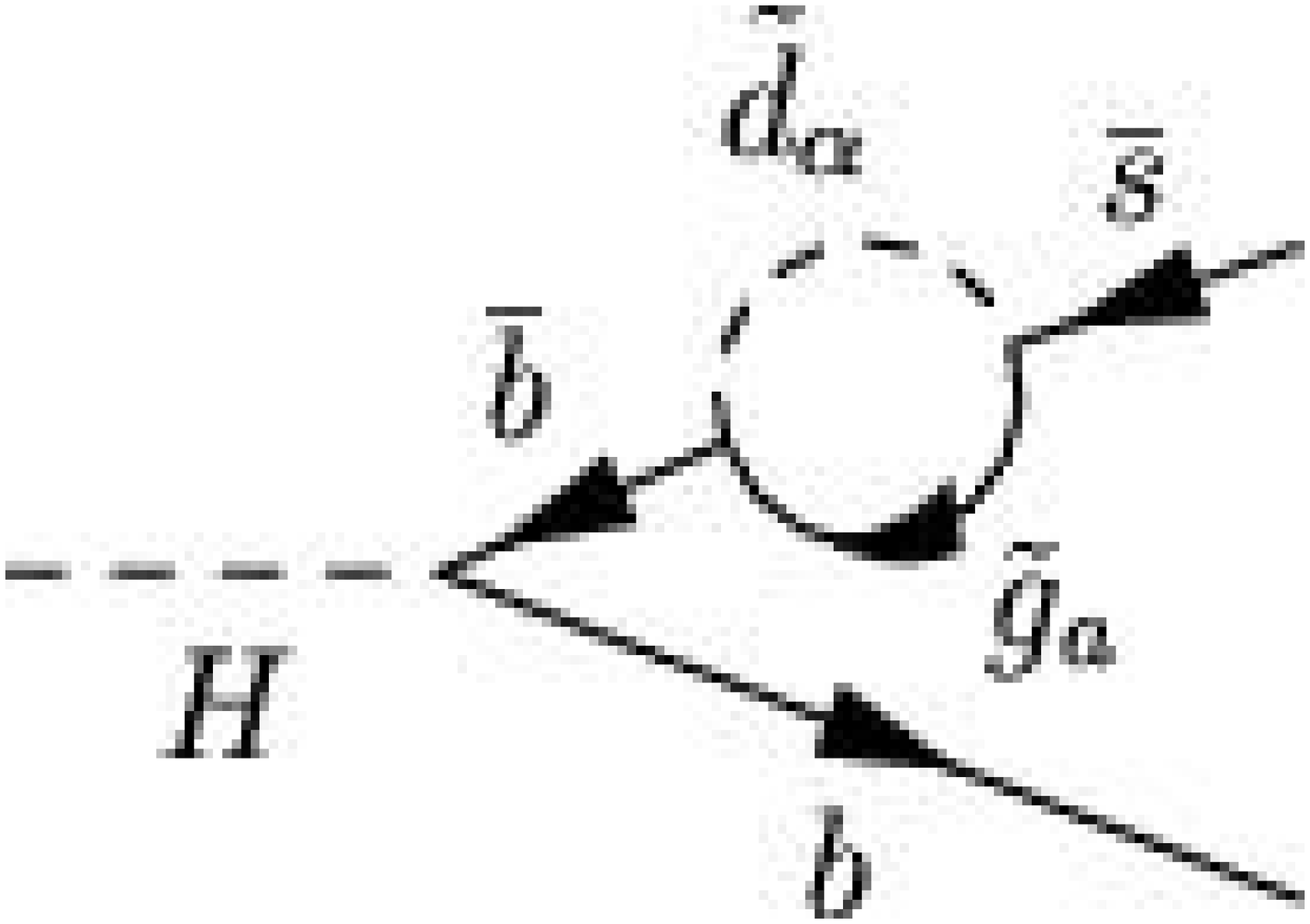,height=1.2in}
\caption{One-loop diagrams from the SUSY-QCD sector for the decay $H \to b \bar s$}
\label{fig.diagrams1}
\end{figure}

For the mixed self-energy 
diagrams it is useful to define the following decomposition,
\begin{eqnarray} 
\Sigma^{bs} (k) &=& k{\hspace{-6pt}\slash} \Sigma_L^{bs} (k^2) P_L + 
 k{\hspace{-6pt}\slash} 
\Sigma_R^{bs} (k^2) P_R + m_b (\Sigma_{Ls}^{bs}(k^2) P_L + 
\Sigma_{Rs}^{bs} (k^2) P_R), \nn \\
\end{eqnarray}
and similarly for $\Sigma^{tc} (k)$ by replacing the quark flavour indices
correspondingly, namely, 
$b \rightarrow t$ and $s \rightarrow c$.

By computing the diagrams in fig.\ref{fig.diagrams1}, 
 we get the following results for the $d$-sector 
form factors in terms of the scalar one-loop integral functions $B_0$, $B_1$, $C_0$, $C_{11}$ and $C_{12}$,
 as defined in refs.~\cite{Hollik},

\begin{eqnarray}
 F_L^{bs} (H_a) & = &- \frac{g_{H_a \tilde d_{\alpha} \tilde d_{\beta}}}{ig} \frac{2 \alpha_s}{3 \pi} 
 \left(m_b R_{3 \alpha}^{(d)} R_{1 \beta}^{(d)*}  (C_{11} - C_{12}) 
 + m_s R_{4 \alpha}^{(d)} R_ {2 \beta}^{(d)*}  C_{12}  \right. \nn \\
&&\left. + M_{ \tilde g} R_{4 \alpha}^{(d)} R_{1 \beta}^{(d)*} C_0 \right) (m_b^2,m_{H_a}^2,m_s^2,M_{\tilde g}^2,M_{\tilde
 d_{\alpha}}^2,M_{\tilde d_{\beta}}^2)
  \nn \\
 && + \frac{G_{H s \bar s}}{i g} \frac{m_b}{m_b^2- m_s^2} \kappa_L^d \left[ m_b (\Sigma_R^{bs} (m_b^2) + 
 \Sigma_{Rs}^{bs} (m_b^2)) + m_s (\Sigma_L^{bs} (m_b^2) + \Sigma_{Ls}^{bs} (m_b^2)) \right] 
\nn \\
 && + \frac{G_{H b \bar b}}{i g} \frac{1}{m_s^2 - m_b^2} \kappa_L^d \left[ m_s^2 \Sigma_L^{bs} (m_s^2) 
 + m_b m_s (\Sigma_{Rs}^{bs} (m_s^2) + \Sigma_R^{bs} (m_s^2))
\right.  \nn \\
&&  \left. + m_b^2 \Sigma_{Ls}^{bs} (m_s^2) \right]   
\label{formfactorLbs}
\end{eqnarray}
\begin{eqnarray}
F_R^{bs} (H_a) & = &- \frac{g_{H_a \tilde d_{\alpha} \tilde d_{\beta}}}{ig} 
\frac{2 \alpha_s}{3 \pi}  \left(m_b R_{4 \alpha}^{(d)} R_{2 \beta}^{(d)*} 
 (C_{11} - C_{12}) 
+ m_s R_{3 \alpha}^{(d)} R_ {1 \beta}^{(d)*}  C_{12} \right. 
\nn \\
&& \left. + M_{ \tilde g} R_{3 \alpha}^{(d)} R_{2 \beta}^{(d)*} C_0 \right) 
(m_b^2,m_{H_a}^2,m_s^2,M_{\tilde g}^2,M_{\tilde d_{\alpha}}^2,M_{\tilde 
d_{\beta}}^2) 
\nn \\
&& + \frac{G_{H s \bar s}}{i g} \frac{m_b}{m_b^2- m_s^2} \kappa_R^d 
\left[ m_b (\Sigma_L^{bs} (m_b^2) + \Sigma_{Ls}^{bs} (m_b^2)) + m_s 
(\Sigma_R^{bs} (m_b^2) + \Sigma_{Rs}^{bs} (m_b^2)) \right]  
\nn \\
&& + \frac{G_{H b \bar b}}{i g} \frac{1}{m_s^2 - m_b^2} \kappa_R^d 
\left[ m_s^2 \Sigma_R^{bs} (m_s^2)
  + m_b m_s (\Sigma_{Ls}^{bs} (m_s^2) + \Sigma_L^{bs} (m_s^2)) \right.  
\nn \\
&&  \left. + m_b^2 \Sigma_{Rs}^{bs} (m_s^2) \right].
\label{formfactorRbs}
\end{eqnarray}
Similarly, by computing the equivalent diagrams for the $tc$ decays, we find
the expresions for the $u$-sector form factors, $F_L^{tc} (H_a)$ and 
$F_R^{tc} (H_a)$, 
 which are identical to the previous results for  
$F_L^{bs} (H_a)$ and  $F_R^{bs} (H_a)$ respectively, but replacing the flavour indices
correspondingly, namely, $b \rightarrow t$, $s \rightarrow c$ and 
$d \rightarrow u$.

In the previous formulas, the $g_{H_a \tilde q_{\alpha} \tilde q_{\beta}}$ are 
the Higgs-squark-squark couplings in the mass eigenstate basis, which are 
collected in the Appendix A. The Higgs-quark-quark couplings are given by,
$G_{H q \bar q} = - \frac{i g m_q}{2 M_W \cos\beta}$, for $q=s,b$, and  
$G_{H q \bar q} = - \frac{i g m_q}{2 M_W \sin\beta}$, for $q=c,t$. The
$\kappa$ factors depend on the particular Higgs decay. These are,  
$\kappa_L^{u} = (\cos\alpha, \sin \alpha, i \cos \beta)$, 
$\kappa_L^{d} = (-\sin\alpha, \cos \alpha, i \sin \beta)$, and 
$\kappa_R^{(u,d)}=\kappa_L^{(u,d)*}$ for the decays of   
$H_a=(h_o,H_o,A_o)$, respectively. 
The last parenthesis in the second 
line of all the form factors refer to the arguments of the 
one-loop $C_0$, $C_{11}$ 
and $C_{12}$ functions. 
Finally,  the one-loop 
contributions from the SUSY-QCD sector to the mixed self-energies 
appearing in the previous equations are the following,
\begin{eqnarray}
     \Sigma_L^{tc,bs} (k^2) &=&  - \frac{2 \alpha_s}{3 \pi} B_1 (k^2,M_{\tilde g}^2,
     M_{\tilde u_{\alpha}, \tilde d_{\alpha}}^2) R_{3 \alpha}^{(u,d)} R_{1 \alpha}^{(u,d)*} \nn \\
     \Sigma_R^{tc,bs} (k^2) &=&  - \frac{2 \alpha_s}{3 \pi} B_1 (k^2,M_{\tilde g}^2,
      M_{\tilde u_{\alpha}, \tilde d_{\alpha}}^2) R_{4 \alpha}^{(u,d)} R_{2 \alpha}^{(u,d)*} \nn \\
     m_{t,b} \Sigma_{Ls}^{tc,bs} (k^2) &=&  - \frac{2 \alpha_s}{3 \pi} M_{\tilde g} 
     B_0 (k^2,M_{\tilde g}^2,M_{\tilde u_{\alpha}, \tilde d_{\alpha}}^2) R_{4 \alpha}^{(u,d)} 
     R_{1 \alpha}^{(u,d)*} \nn \\
     m_{t,b} \Sigma_{Rs}^{tc,bs} (k^2) &=&  - \frac{2 \alpha_s}{3 \pi}  M_{\tilde g} 
     B_0 (k^2,M_{\tilde g}^2,M_{\tilde u_{\alpha}, \tilde d_{\alpha}}^2) R_{3 \alpha}^{(u,d)} R_{2 \alpha}^{(u,d)*}
\label{mixed_self_energies} 
\end{eqnarray}
The expressions for the form factors involved in the $H_a \to c \bar t, \, t \bar c$ decays
have been checked to be in agreement with the previous results
obtained in~\cite{sola} for the $t \to c H_a$ decays.

 We next present the results of the partial decay widths for the decays 
$H_a \to t\bar c, c\bar t$ where
$H_a = H_o, A_o$ and $H_a \to b\bar s, s\bar b$ 
where $H_a = h_o, H_o, A_o$, in terms of the above form factors.
Since we are assuming that the final states $q \bar q'$ and  $q' \bar q$ 
are not experimentally distinguishable, 
the final results for the total partial widths are got by adding the 
two corresponding partial widths, and this will be denoted from now 
on by $\Gamma(H \to q \bar q' + q' \bar q)$. These results are as follows,

\begin{eqnarray}
\Gamma (H_a \to b\bar s+ s\bar b) &=& \frac{2 g^2}{16 \pi m_{H_a}} \lambda^{\frac{1}{2}} \left(1,\frac{m_s^2}{m_{H_a}^2}, \frac{m_b^2}{m_{H_a}^2} \right) 
\times \nn \\
&& \left[3 (m_{H_a}^2 - m_s^2 - m_b^2) 
(F_L^{bs}(H_a) F_L^{bs*}(H_a) 
+ F_R^{bs}(H_a) F_R^{bs*}(H_a)) \right. \nn \\
&& \left. - 6 m_b m_s (F_L^{bs}(H_a) F_R^{bs*}(H_a) 
+ F_R^{bs}(H_a) F_L^{bs*}(H_a)) \right] \nn \\
\end{eqnarray}
%\vspace{1cm}
\begin{eqnarray}
\Gamma (H_a \to t\bar c+ c\bar t) &=& \frac{2 g^2}{16 \pi m_{H_a}} \lambda^{\frac{1}{2}} \left( 1,\frac{m_c^2} 
{m_{H_a}^2}, \frac{m_t^2}{m_{H_a}^2} \right) \times \nn \\
&& \left[3 (m_{H_a}^2 - m_c^2 - m_t^2)  (F_L^{tc} (H_a)F_L^{tc*}(H_a)  
+ F_R^{tc}(H_a) F_R^{tc*}(H_a)) \right. \nn \\
&&\left. - 6 m_t m_c (F_L^{tc}(H_a) F_R^{tc*}(H_a) + F_R^{tc}(H_a) F_L^{tc*}
(H_a)) \right]  
\end{eqnarray}
where $\lambda^{\frac{1}{2}} \left( 1, x^2, y^2 \right) = 
\sqrt{ \left[ 1 - (x + y )^2 \right] \left[ 1 - (x - y)^2 \right]}$.
Notice again that, due to phase space, the lightest Higgs boson $h_o$ 
cannot decay into $t \bar c$ or $ c \bar t$. 

Some comments are in order. First, we have checked explicitly that the 
previous results for the form factors are finite, as expected.  
Second, we can see that 
the parameter $\lambda$, which 
characterizes the FC, does not appear explicitly
in eqs.(\ref{formfactorLbs}) and (\ref{formfactorRbs}), but it appears 
implicitly.
As we have already explained, the effect of FC is due to misalignment 
between the quark and squark mass matrices and is parametrized through 
non-diagonal terms in the squark squared mass matrices containing the 
parameter $\lambda$. Therefore, the dependence on 
$\lambda$ is hidden in the respective rotation matrices 
used to diagonalize the squark squared mass matrices, $R_{\alpha \beta}^{(u)}$ 
and $R_{\alpha \beta}^{(d)}$, and obviously in the physical squark mass 
values, which appear as arguments of the scalar one-loop integral functions 
obtained in the one-loop calculation.

On the other hand, the MSSM parameters $\tan \beta$, 
$\mu$ and $M_{\tilde g}$ will be crucial for our phenomenological 
analysis of section~\ref{chap.phenomenology}, where we will study numerically 
the size of these loop induced FCHD as a function of these MSSM parameters. 
The dependence on $\mu$ and $\tan \beta$ in the form factors, and hence, 
in the decay widths, is complicated and is hidden inside couplings and 
inside the squark mass matrices.
$M_{\tilde g}$ appears as an argument of the scalar one-loop integral 
functions, and also it appears explicitly as a multiplicative factor in both 
the vertex $C_0$ contributions and the scalar parts of the mixed 
self-energy diagrams. The global dependence on the angle $\beta$ 
is more complicated, as it emerges in most of the terms. 
The explicit $\beta$ dependence through  $LR$ flavour 
preserving squark mixing, namely, $m_{b,s} (A_{b,s}- \mu \tan \beta)$ and 
$m_{t,c} (A_{t,c}- \mu \cot \beta)$, shows that the contribution 
from these mixing terms in the $down$ sector grows with $\tan \beta$ and 
its dependence on $A_{b,s}$, for large values of $\tan \beta$, is expected to be mild, while 
in the $up$ sector, 
it is less $\tan \beta$ dependent for $\tan \beta > 1$ and the 
trilinear term, $A_{t,c}$, can acquire more importance.

At the end, one crucial parameter for our complete analysis will be $\mu$, 
which appears in the squark mass matrix term in the way exposed above, but 
also, directly in the Higgs-squark-squark couplings, as can be seen in 
the Appendix A. As we will show in the next sections, 
large values of $\mu$ can produce sizable contributions to FCHD. 

All these complicated dependences with the different MSSM parameters will be 
clarified in the analysis performed in the next section, and the analytical 
behaviour will be explained in more detail in the large SUSY mass limit 
performed in 
section~\ref{chap.decoupling}, where we study the behaviour of these FC decay 
processes in the scenario where a very heavy supersymmetric spectrum is 
considered.

%%%%%%%%%%%%%%%%%%%%%%%%%%%%%%%%%%%%%%%%%%%%%%%%%%%%%%%%%%%%%%%%%%%%%%%%%%%%
\section{Numerical Analysis of the FCHD rates}
\label{chap.phenomenology}

In this section  we estimate numerically the size of the 
loop induced FCHD as a function of the MSSM parameters and 
the $\lambda$ parameter. 
 The MSSM parameters needed to fully characterize and evaluate the 
$\Gamma (H_a \to b \bar s + s \bar b)$ and 
$\Gamma (H_a \to t \bar c + c \bar t)$ partial widths, 
in our simplified scenario, consist of the following six parameters: 
$m_A$, $\tan \beta$, $\mu$, $M_{\tilde g}$, $M_o$, and A, where we have 
chosen, for simplicity, $M_o$ as a common value for the soft 
squark breaking mass parameters, 
$M_o = M_{\tilde Q,q} = M_{\tilde U,(c,t)} = M_{\tilde D,(s,b)}$,
 and all the various trilinear parameters to be equal $A_t=A_b=A_c=A_s=A$. 
 These parameters will be varied over a broad range, subject only to our 
 requirement that all 
 the squark masses be heavier than 150 $GeV$, 
 and that the gluino mass be heavier
 than $200 \, GeV$. 
 On the other hand, the extra 
parameter, $\lambda$, which is the only one measuring  the FC strength, 
will be varied in the range $0 \leq \lambda \leq 1$, and only values leading
to $M_{\tilde q} > 150 \, GeV$ will be considered. 
 
 In order to estimate the FCHD rates over the MSSM parameter space and 
 the
  $\lambda$ interval chosen, and with the motivation in mind of finding the 
 region of the 
 MSSM parameter space where these rates are maximal, we will proceed as follows.
 We will first fix the $\lambda$ parameter to one concrete value in the 
 range $0 \leq \lambda \leq 1$ and vary the rest of parameters over the whole
 allowed range in order 
 to detect such maximizing region. Next, we will fix all the MSSM parameters to 
 some specific values belonging to this region and study 
 the  widths and 
 branching ratios as a function of the $\lambda$ parameter. 
 This will give us finally the maximum size of these observables as a 
 function of $\lambda$ as well, and if any planned experiment is able 
 to observe and measure these FCHD rates with good accuracy, 
 their experimental value  could serve to extract either the 
 preferred $\lambda$ value or,  
 in the worst case, an upper bound on it. 
%\newpage

\begin{center}
\begin{figure}[t]
\vspace{-.80cm}
\hspace{-1.0cm}
\epsfig{file=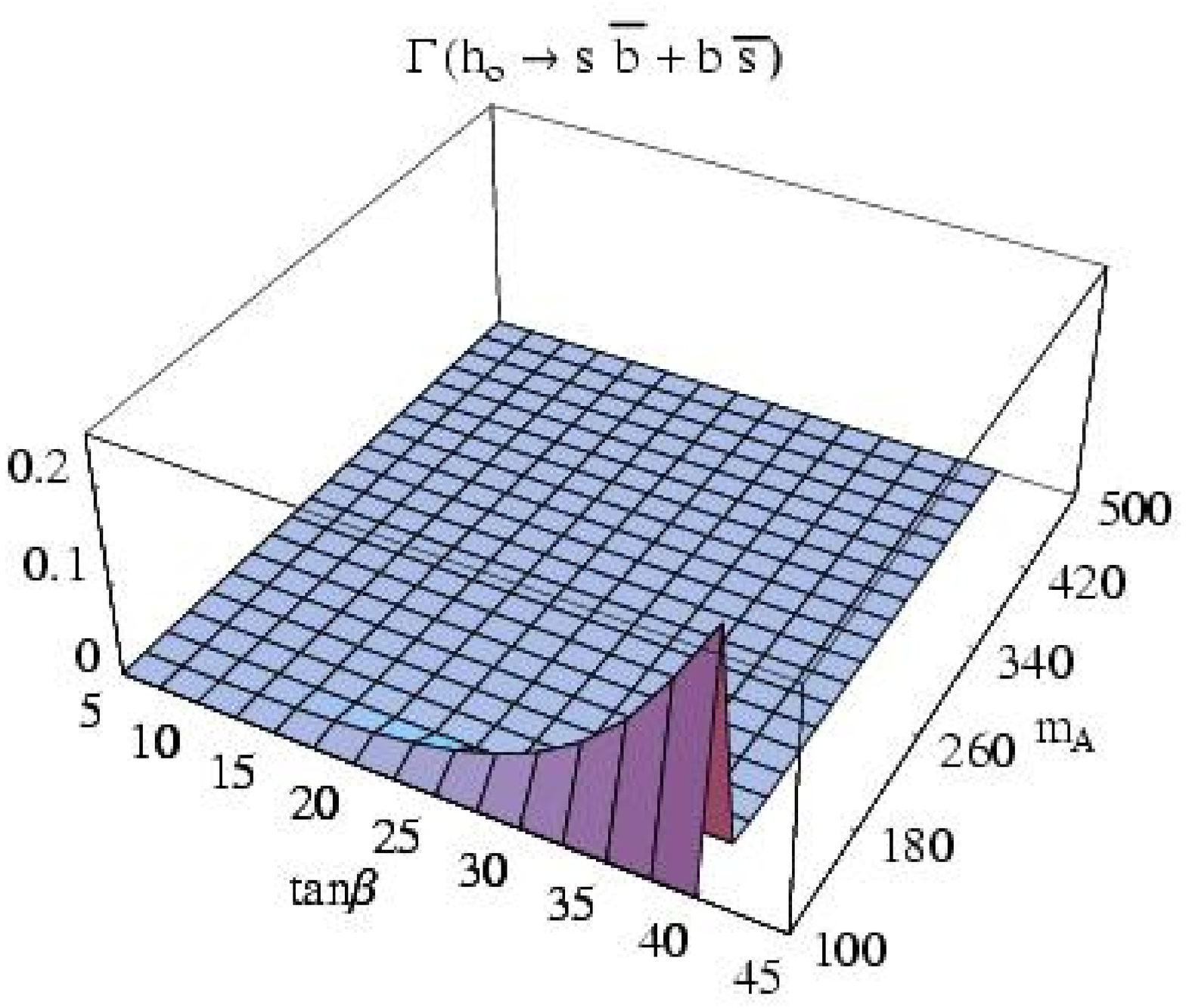,height=2.25in}
\epsfig{file=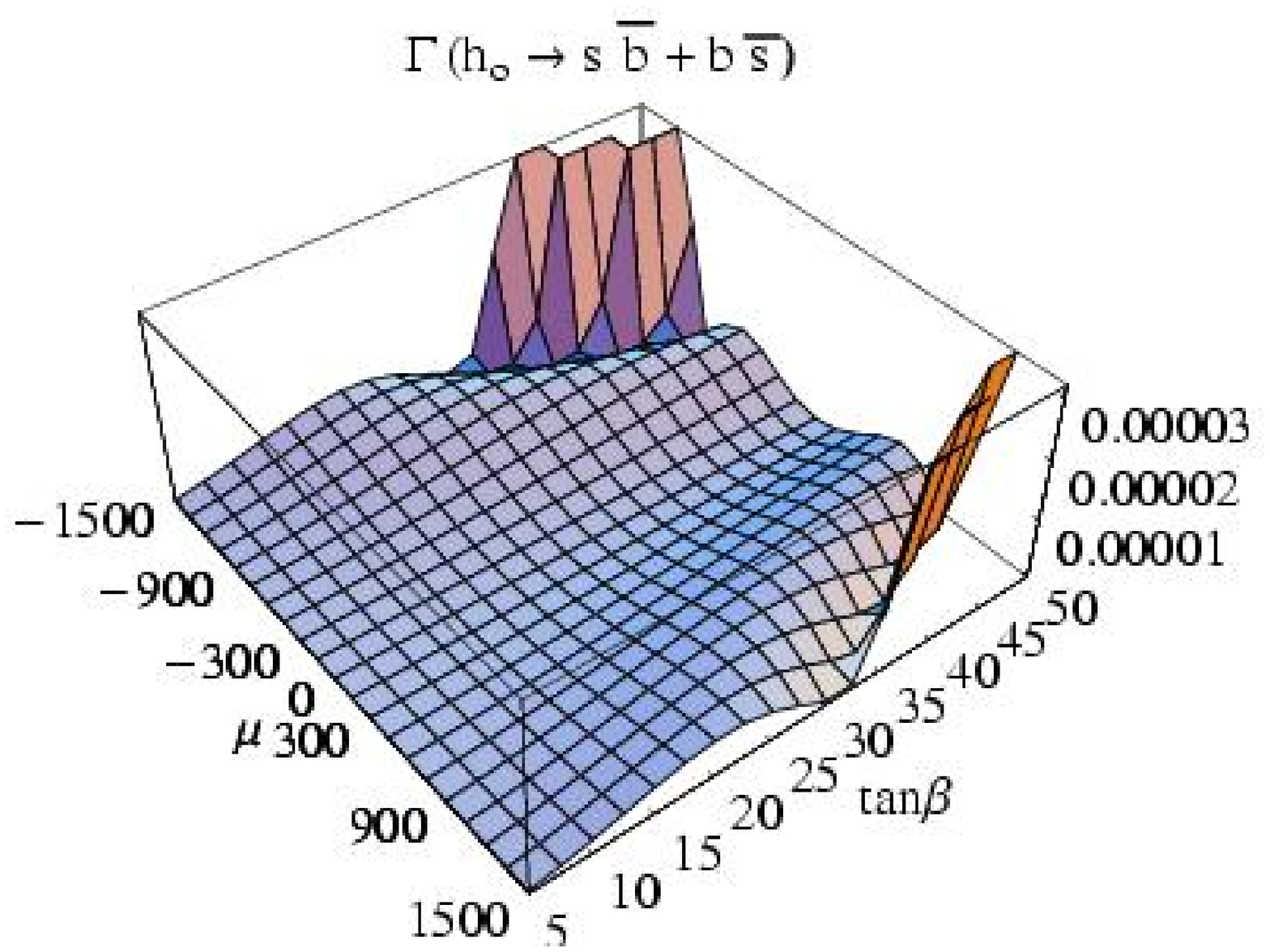,height=2.25in}
\vspace*{1cm}
\vspace{-1.35cm}
\hspace{-1.0cm}
\epsfig{file=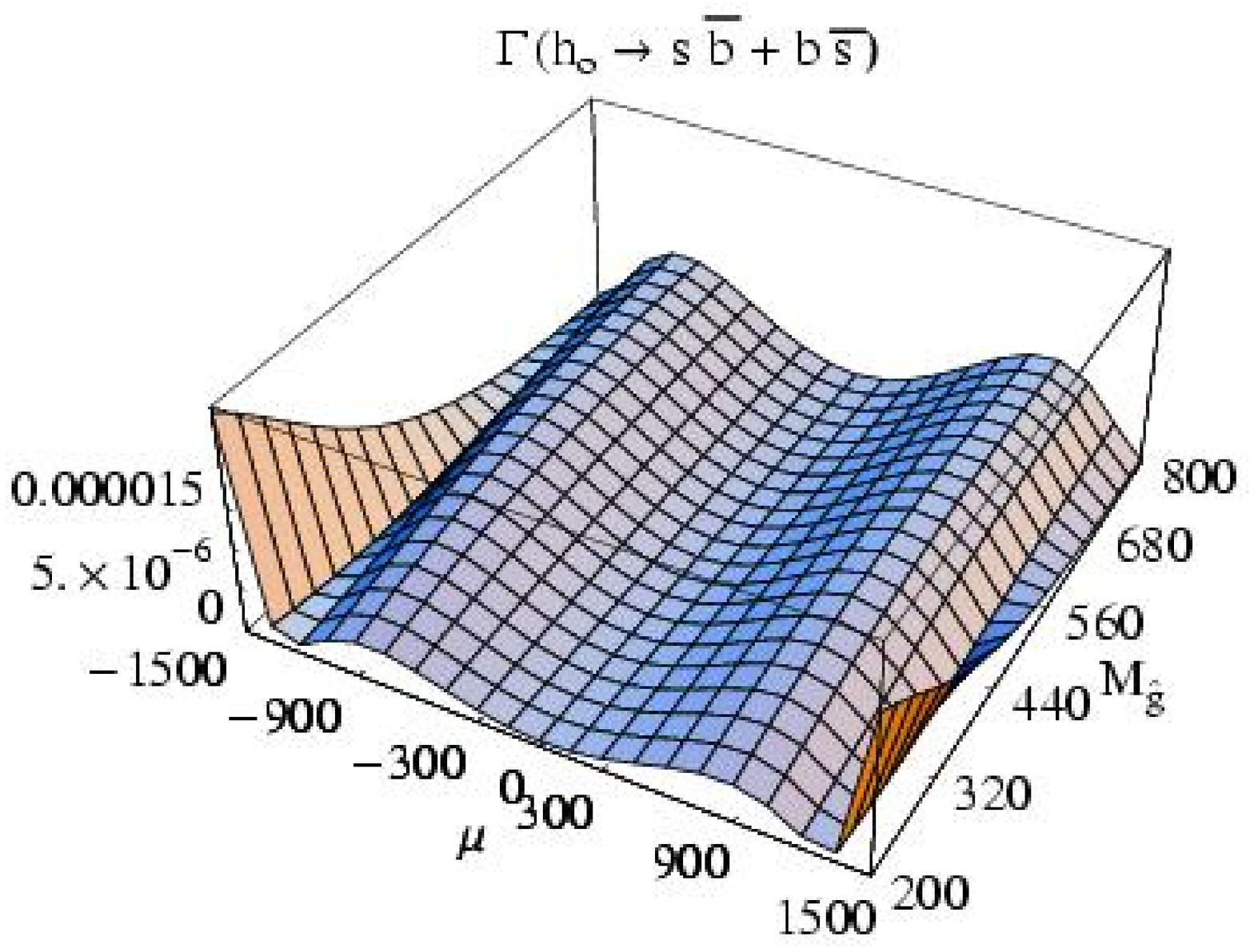,height=2.25in}
\epsfig{file=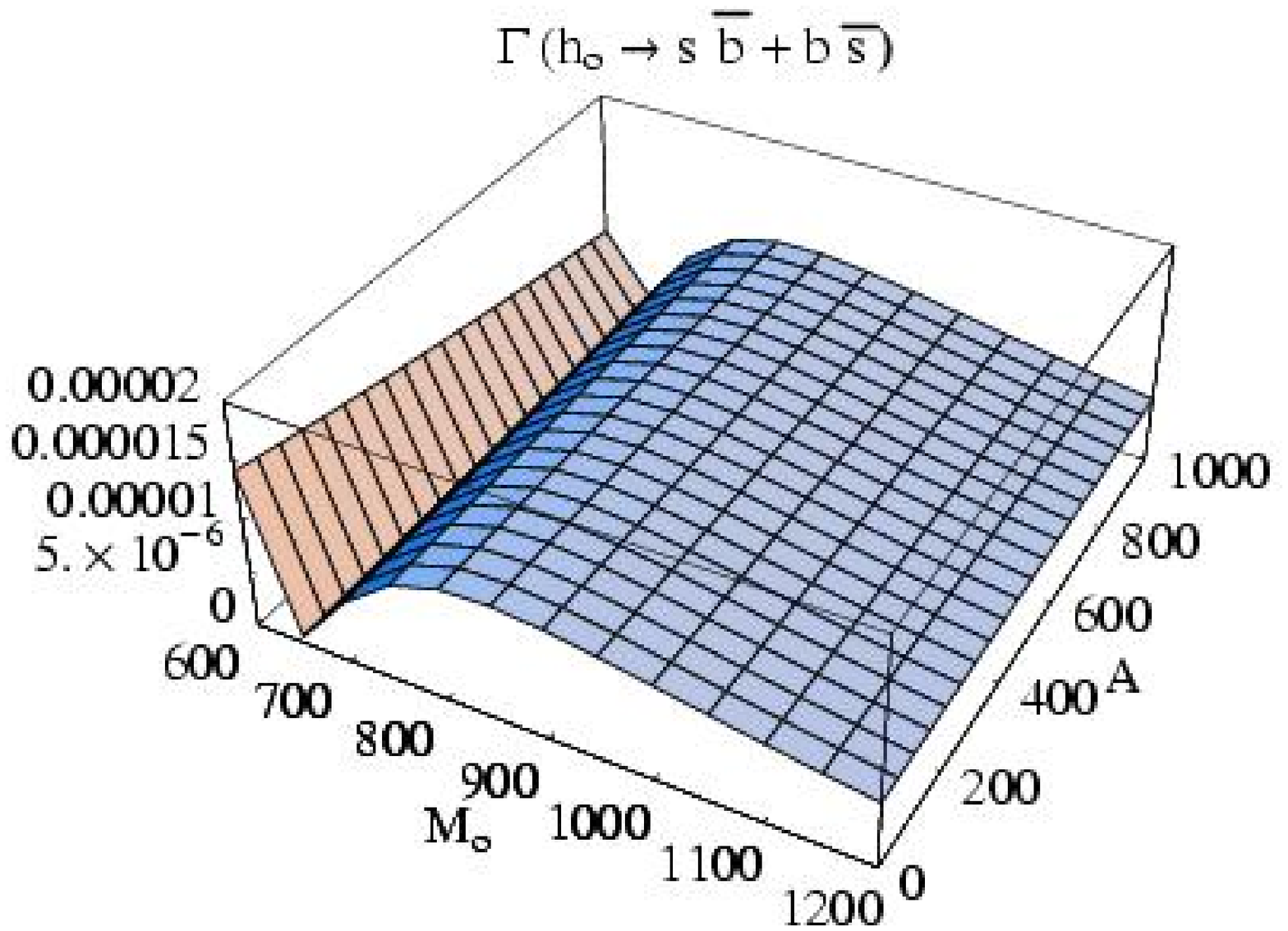,height=2.15in}
\caption{$\Gamma (h_o \to b \bar s + s \bar b)$ in $GeV$ as a function
of ($\tan \beta$, $m_A$ ($GeV$)) 
{\bf (a)}, ($\mu$ ($GeV$), $\tan \beta$) {\bf (b)}, ($\mu$ ($GeV$), $M_{\tilde g}$ ($GeV$)) 
{\bf (c)} and 
($M_o$ ($GeV$), $A$ ($GeV$)) {\bf (d)}. The regions of the parameter space not plotted are the
 ones that 
give non allowed values for the squark masses ($M_{\tilde q} < 150 \, GeV$).
The values of the different MSSM parameters that have to be fixed in each plot have been chosen 
correspondingly to be: $\mu=1500 \, GeV$, $M_o=600 \, GeV$, $M_{\tilde g}= 300 \, GeV$, 
$A = 200 \, GeV$, 
$m_A = 250 \ GeV$, $\tan \beta = 35$ and $\lambda = 0.5$.}
\label{hbs_1}
\end{figure}
\begin{figure}[t]
\vspace{-0.75cm}
\hspace{-1.0cm}
\epsfig{file=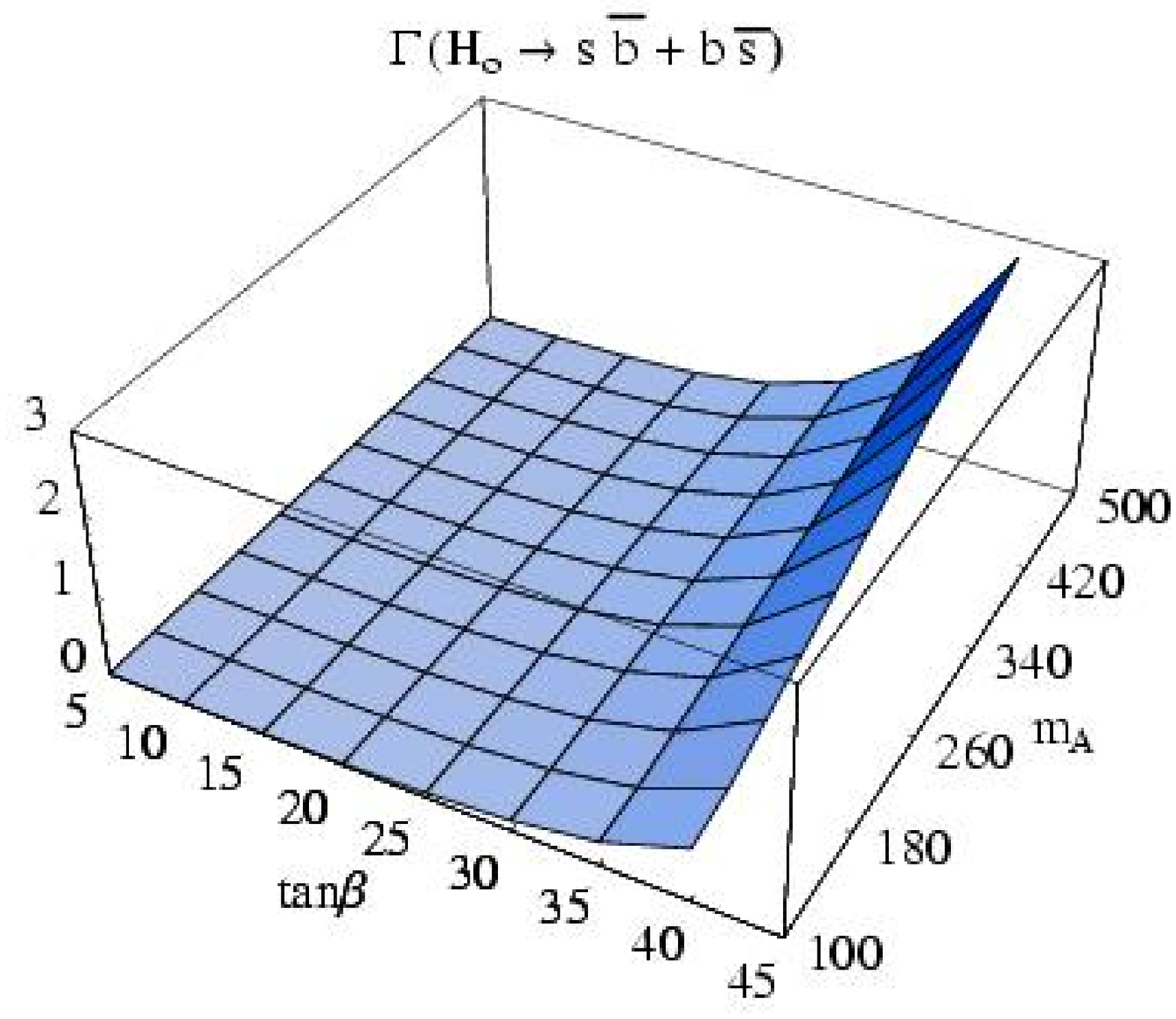,height=2.5in}
\epsfig{file=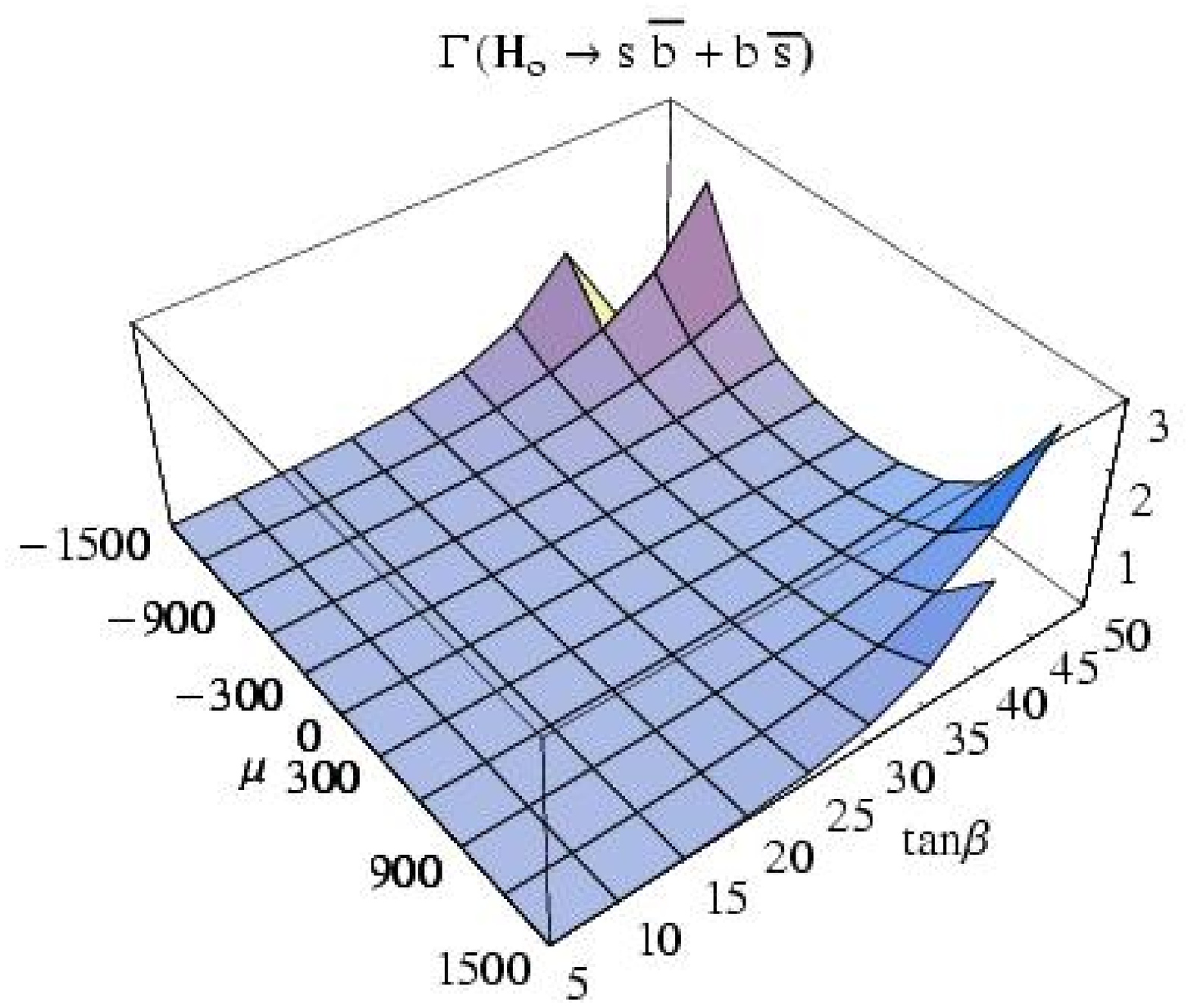,height=2.5in}\\
\vspace*{0.75cm}
\vspace{-1.35cm}
\hspace{-1.0cm}
\epsfig{file=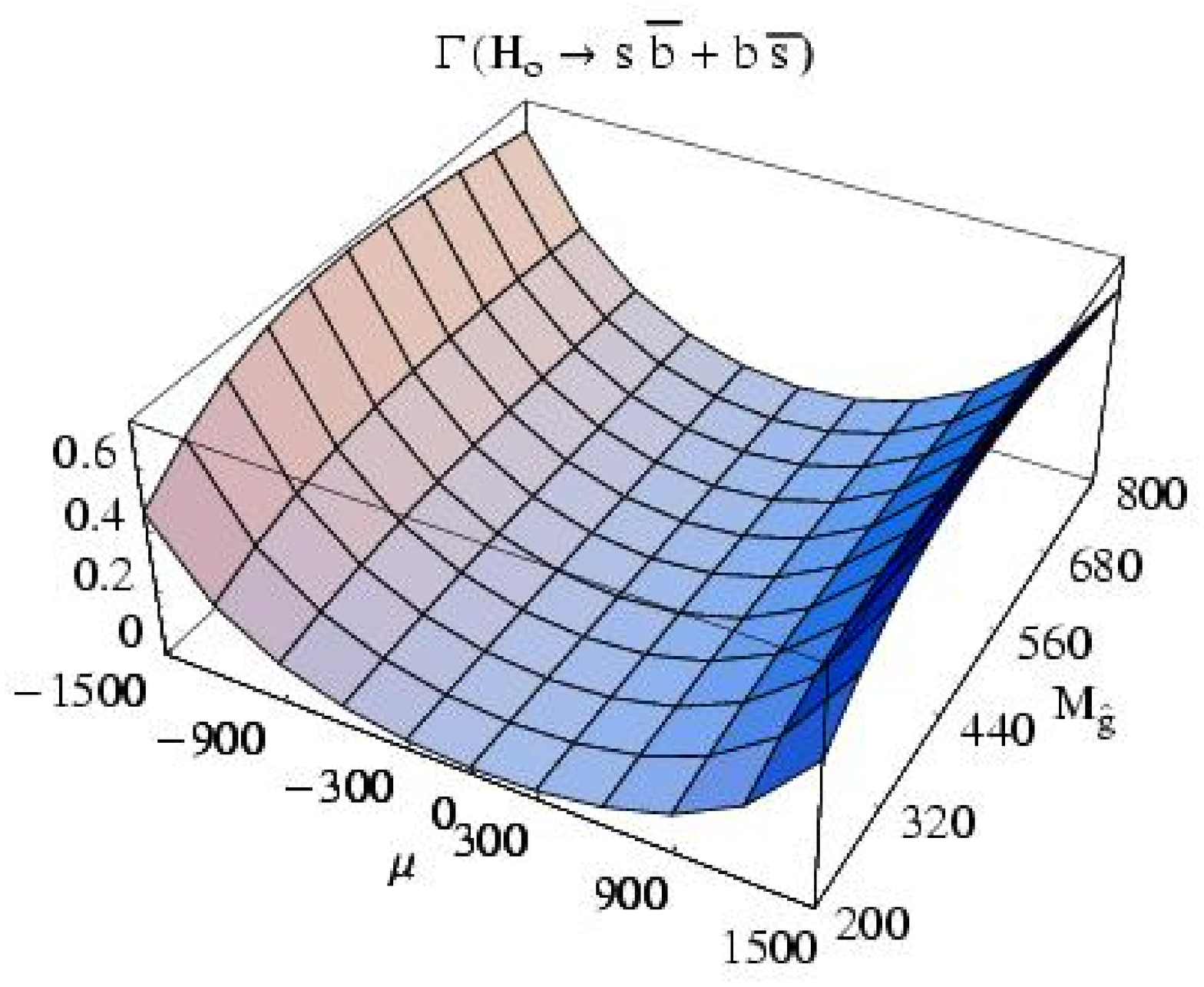,height=2.5in}
\epsfig{file=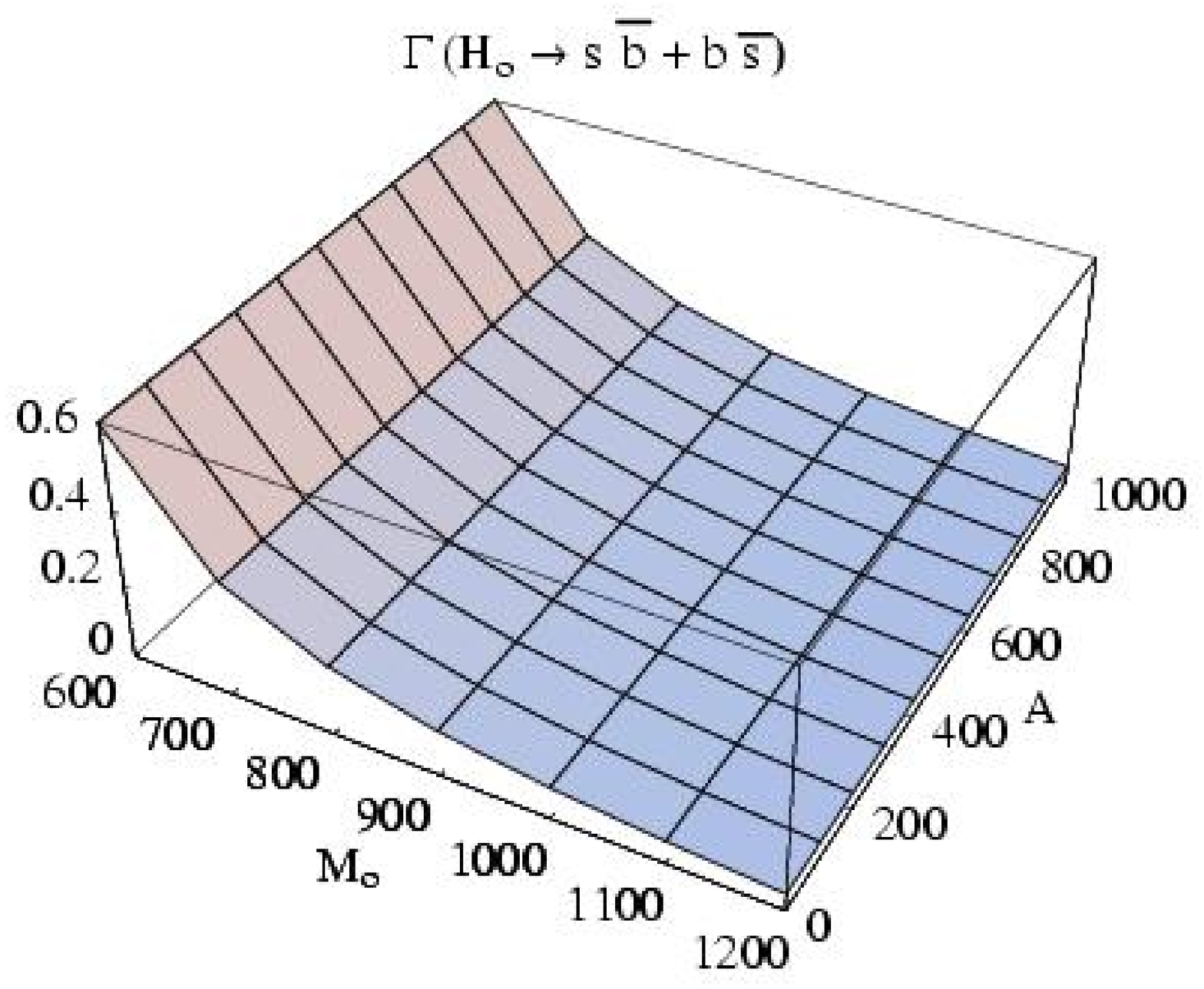,height=2.5in}\vspace{0.4em}
\caption{Same as in fig.\ref{hbs_1} but for $\Gamma (H_o \to b \bar s + s \bar b)$.}
\label{hbs_2}
\end{figure}
\begin{figure}[h]
\vspace{-0.75cm}
\hspace{-1.0cm}
\epsfig{file=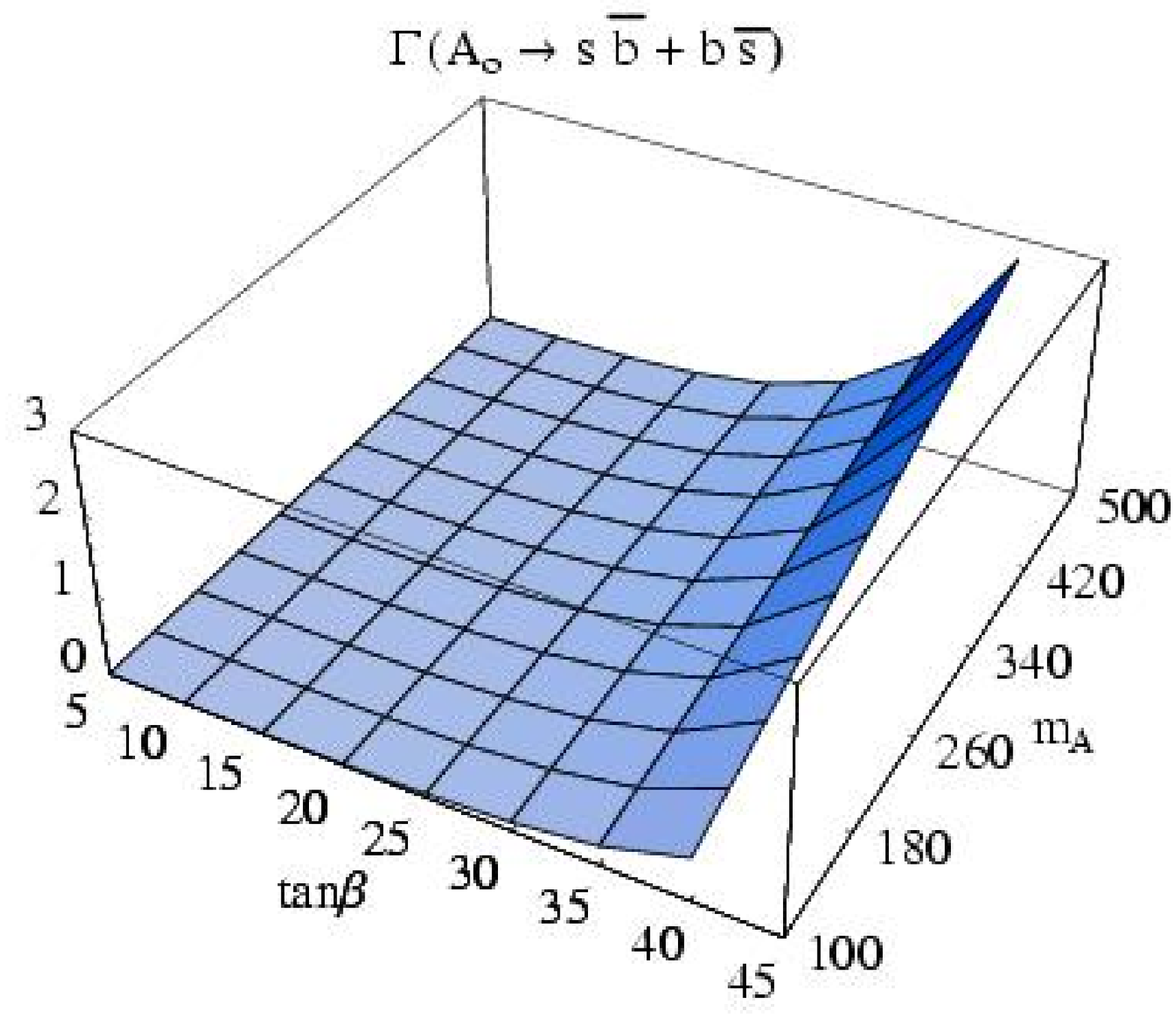,height=2.5in}
\epsfig{file=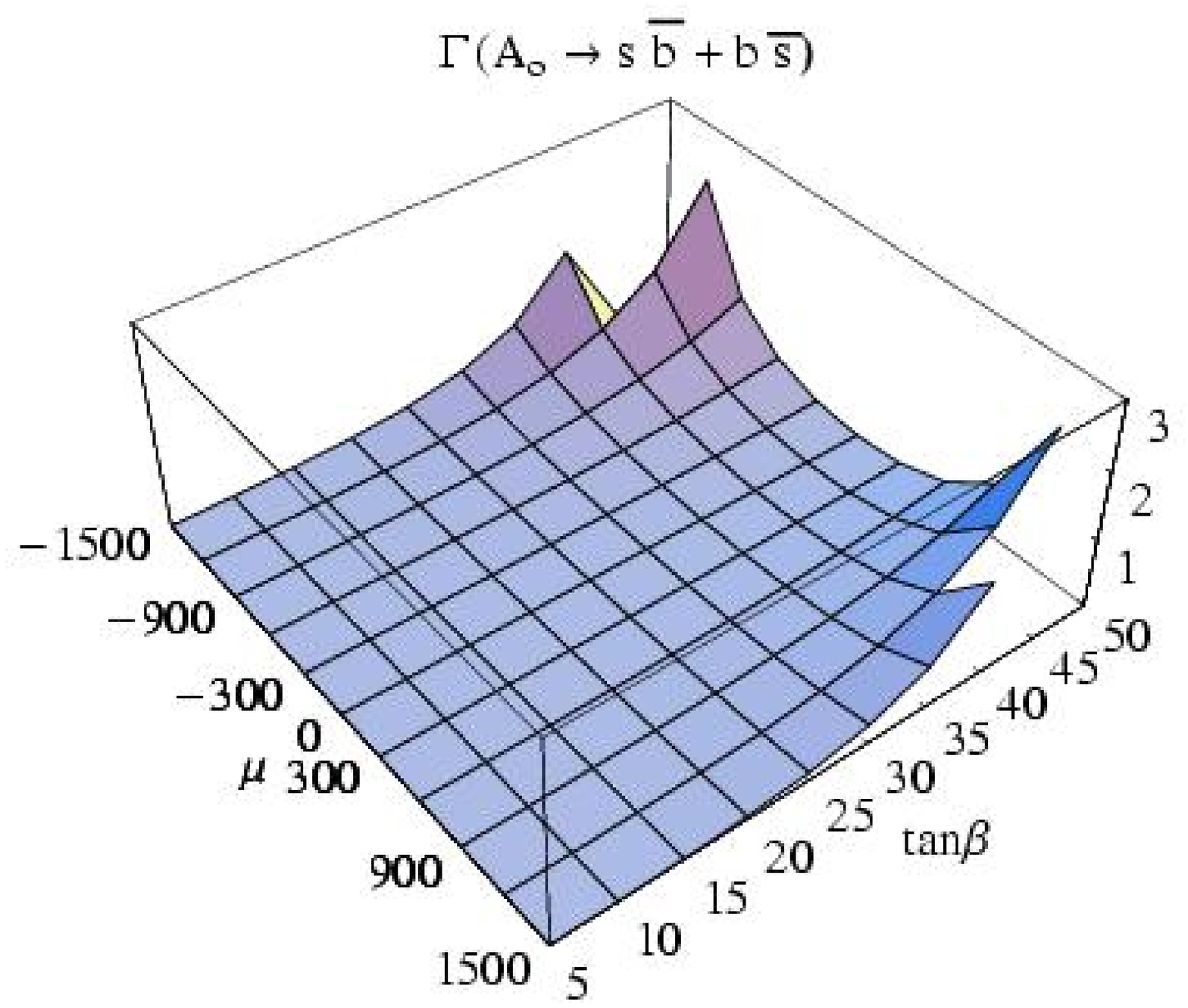,height=2.5in}\\
\vspace*{0.75cm}
\vspace{-1.35cm}
\hspace{-1.0cm}
\epsfig{file=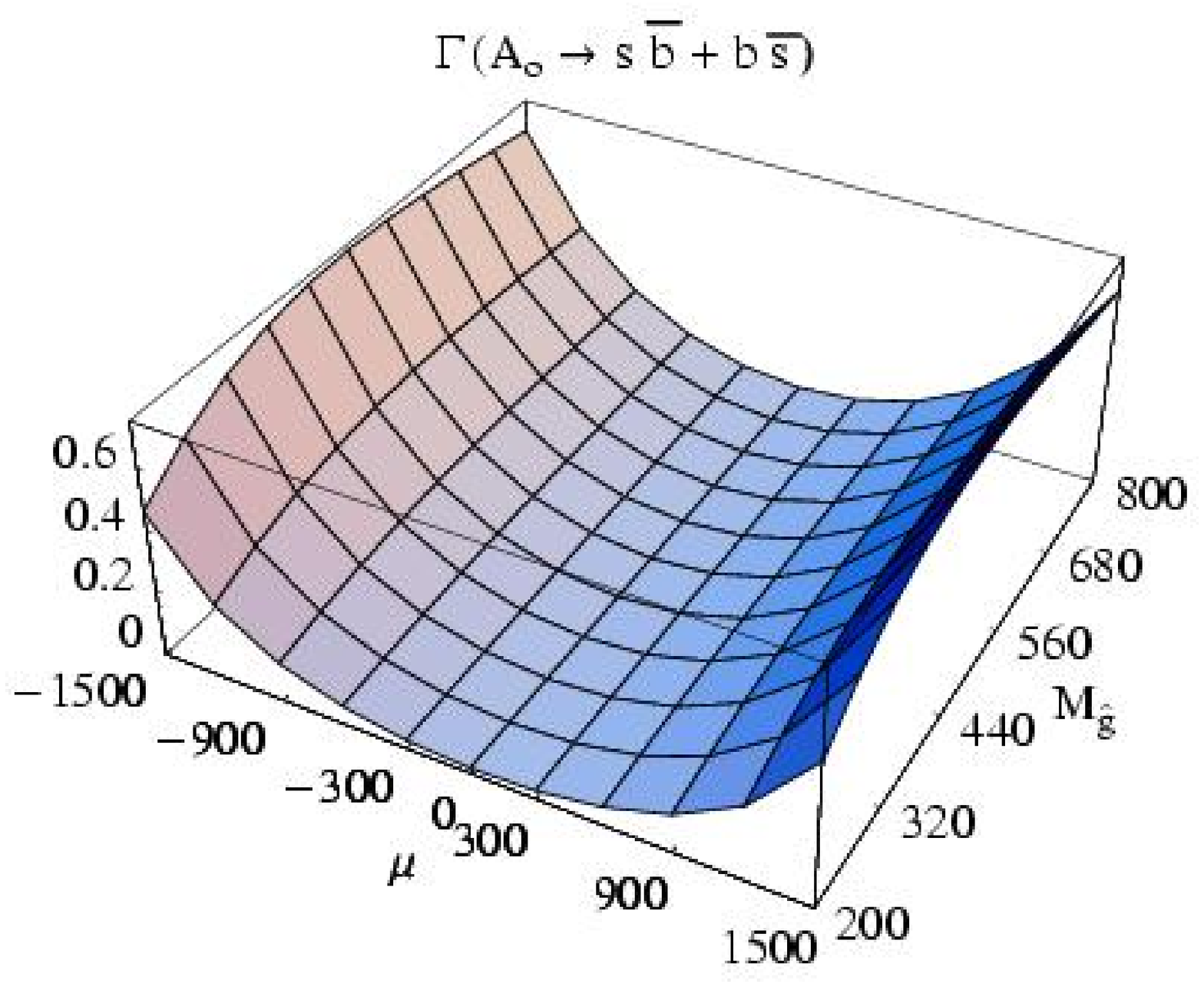,height=2.5in}
\epsfig{file=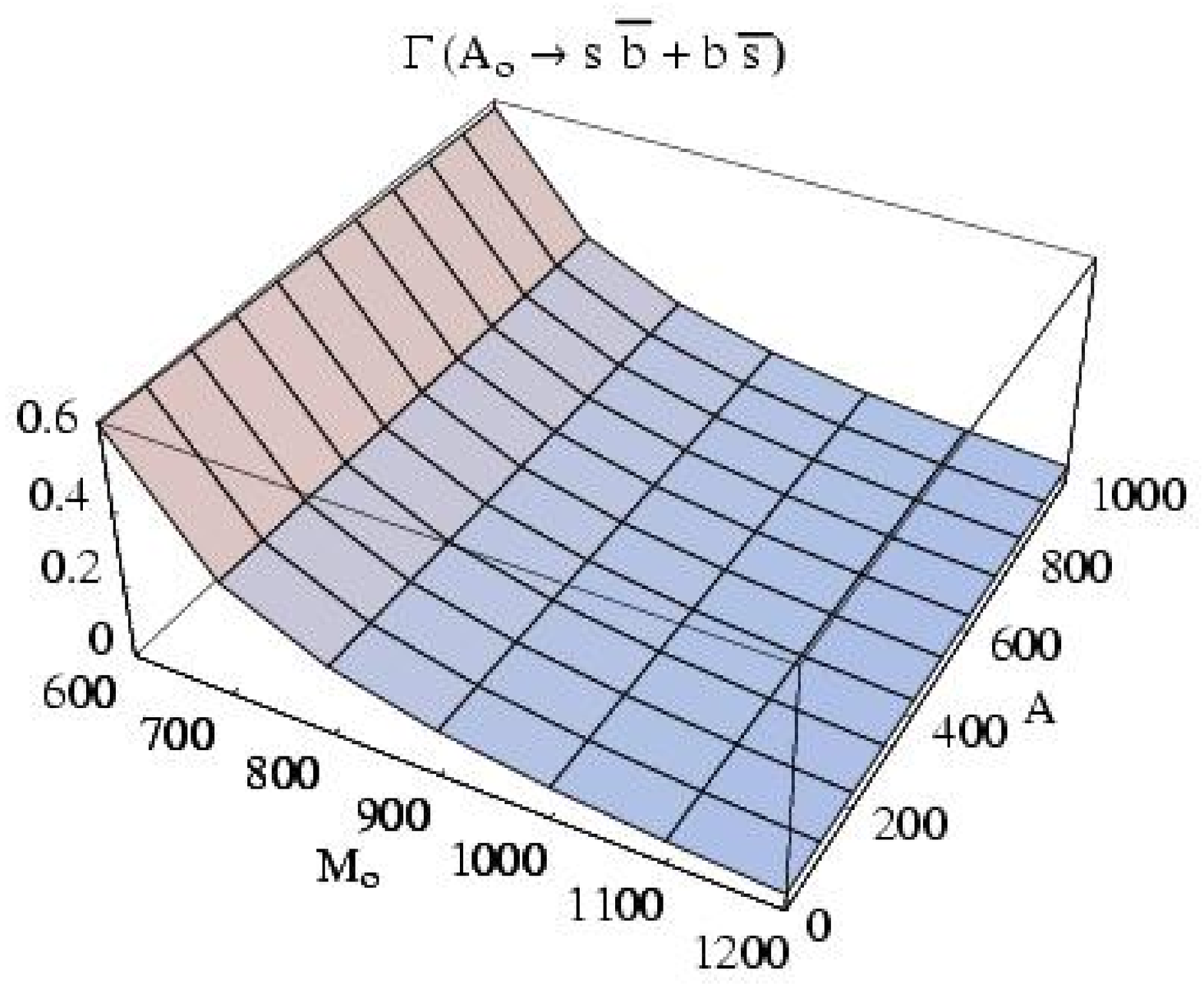,height=2.5in}\vspace{0.4em}
\caption{Same as in fig.\ref{hbs_1} but for $\Gamma (A_o \to b \bar s + s \bar b)$.}
\label{hbs_3}
\end{figure}
\begin{figure}[h]
\vspace{-0.75cm}
\hspace{-1.0cm}
\epsfig{file=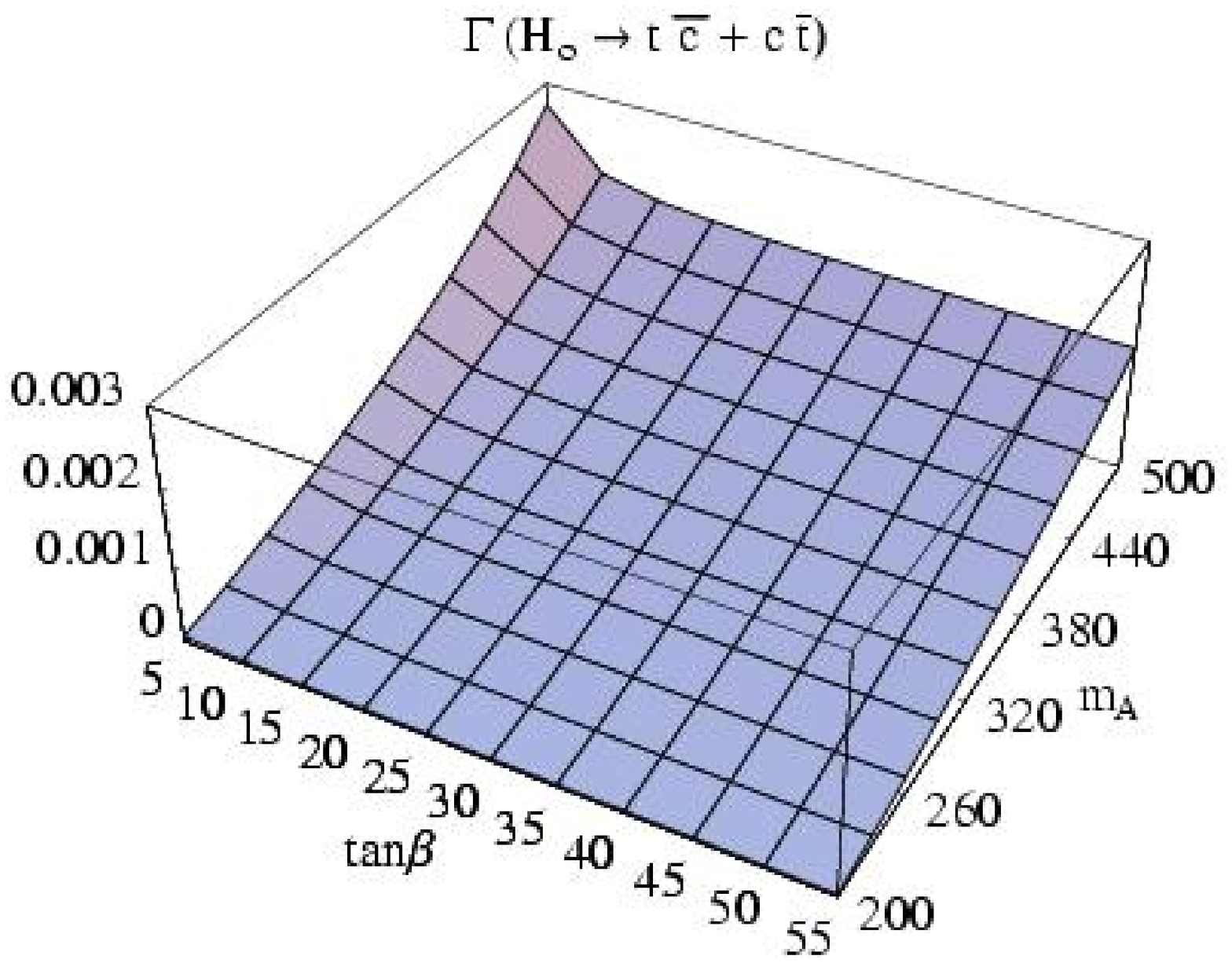,height=2.5in}
\epsfig{file=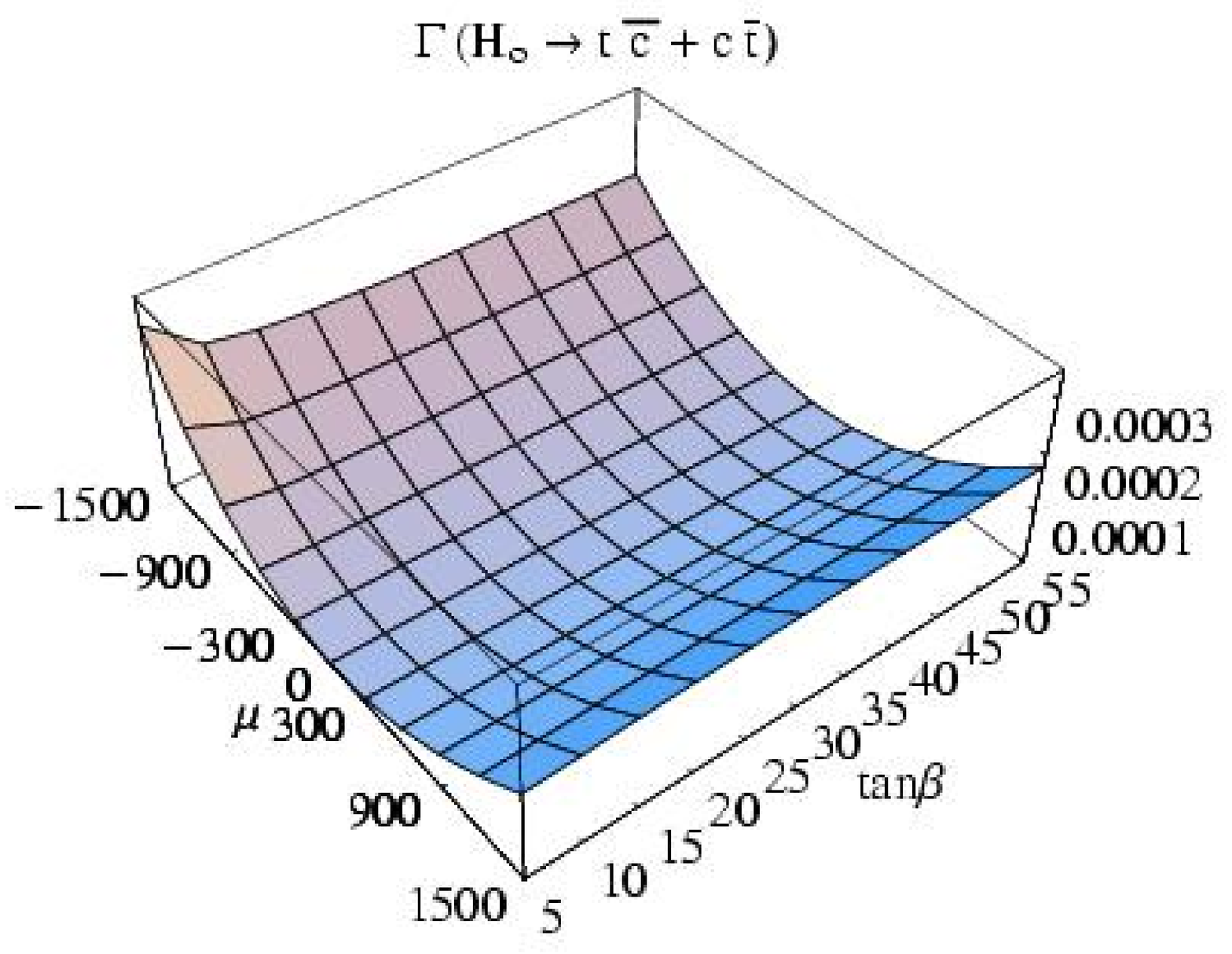,height=2.5in}\\
\vspace*{0.75cm}
\vspace{-1.35cm}
\hspace{-1.0cm}
\epsfig{file=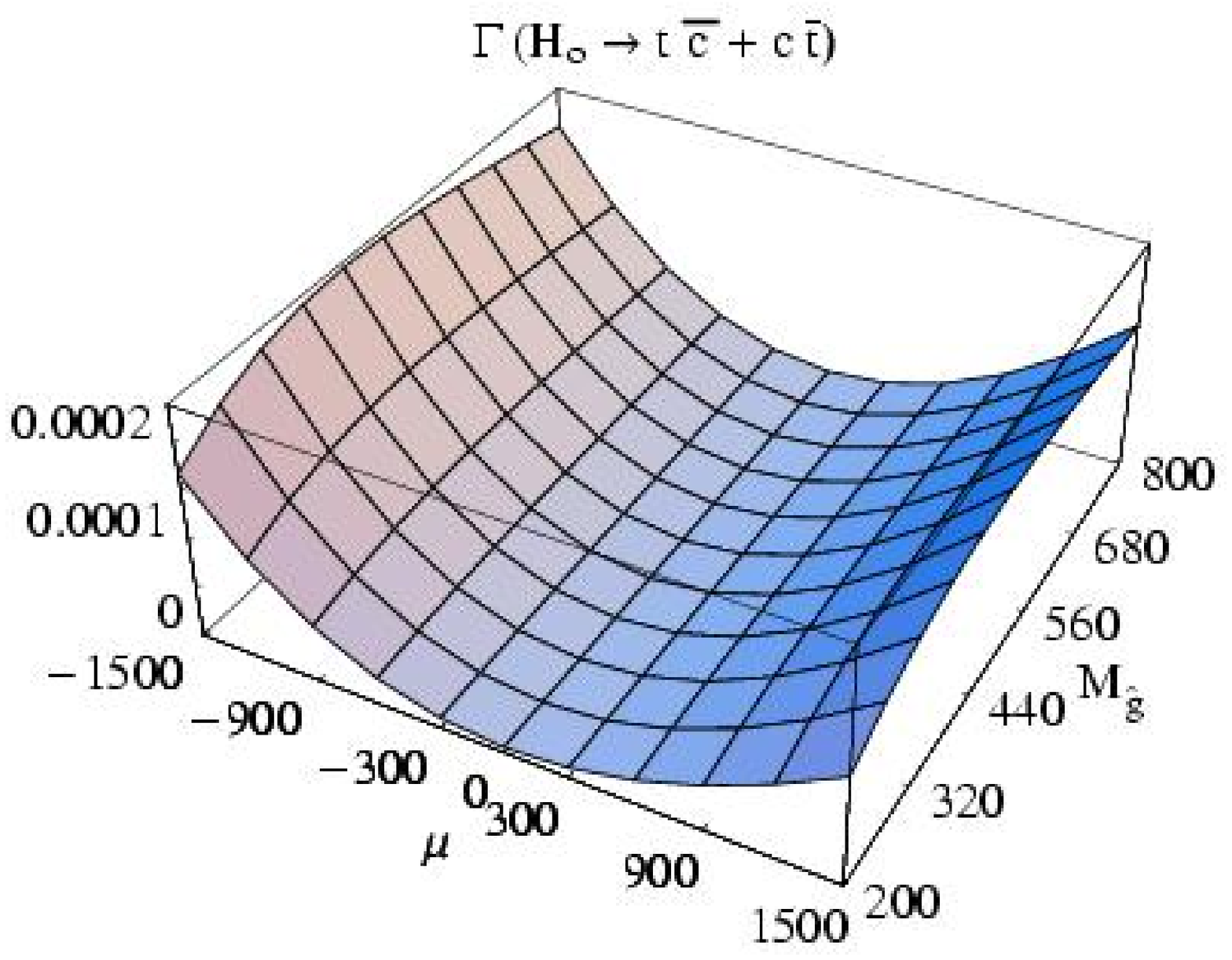,height=2.5in}
\epsfig{file=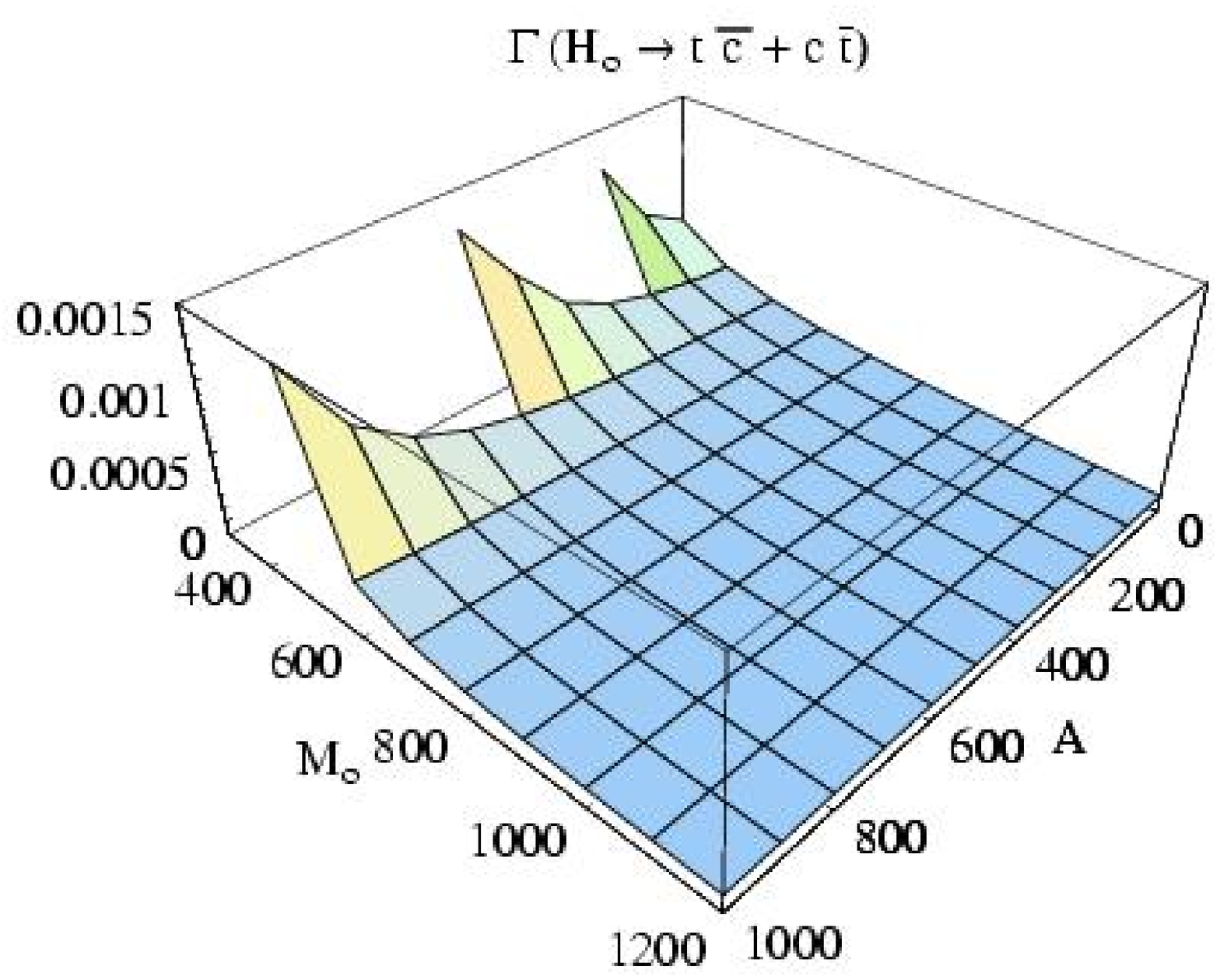,height=2.5in}\vspace{0.4em}
\caption{$\Gamma (H_o \to t \bar c + c \bar t)$ in $GeV$ as a function
of ($\tan \beta$, $m_A$ ($GeV$)) {\bf (a)}, ($\mu$ (GeV), $\tan \beta$) {\bf (b)}, ($\mu$ ($GeV$), $M_{\tilde g}$ 
($GeV$)) {\bf (c)} and ($M_o$ ($GeV$), $A$ ($GeV$)) {\bf (d)}. The regions of the parameter space
 not plotted are the ones that give non allowed values for the squark masses 
 ($M_{\tilde q} < 150 \, GeV$).
The values of the different MSSM parameters that have to be fixed in each plot have been chosen 
correspondingly to be: $\mu = - 1500 \, GeV$, $M_o = 600 \, GeV$, $M_{\tilde g}= 500 \, GeV$, 
$A = 800 \, GeV$, $m_A = 250 \, GeV$ and $\tan \beta = 10$.}
\label{htc_2}
\end{figure}
\begin{figure}[h]
\vspace{-0.75cm}
\hspace{-1.0cm}
\epsfig{file=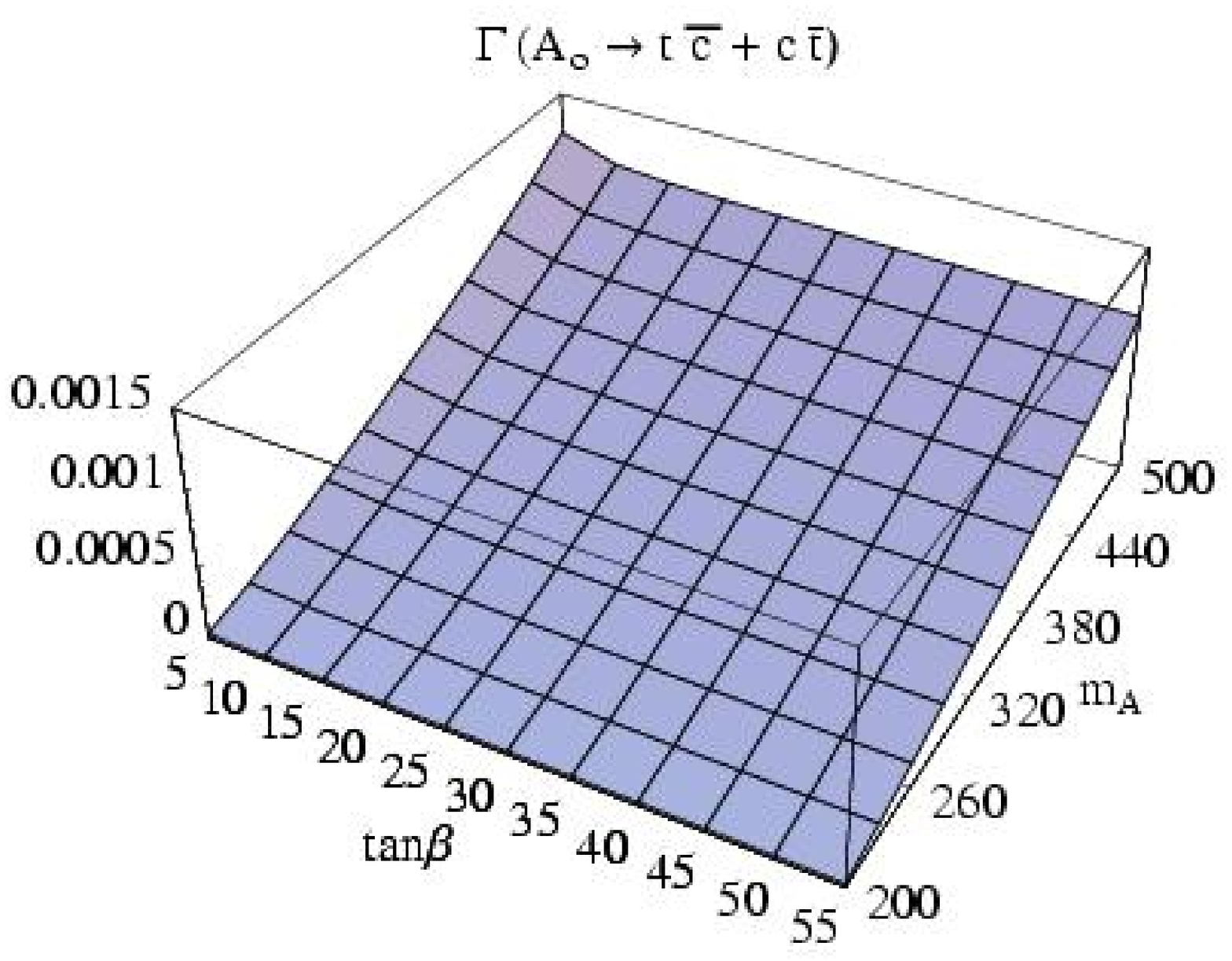,height=2.5in}
\epsfig{file=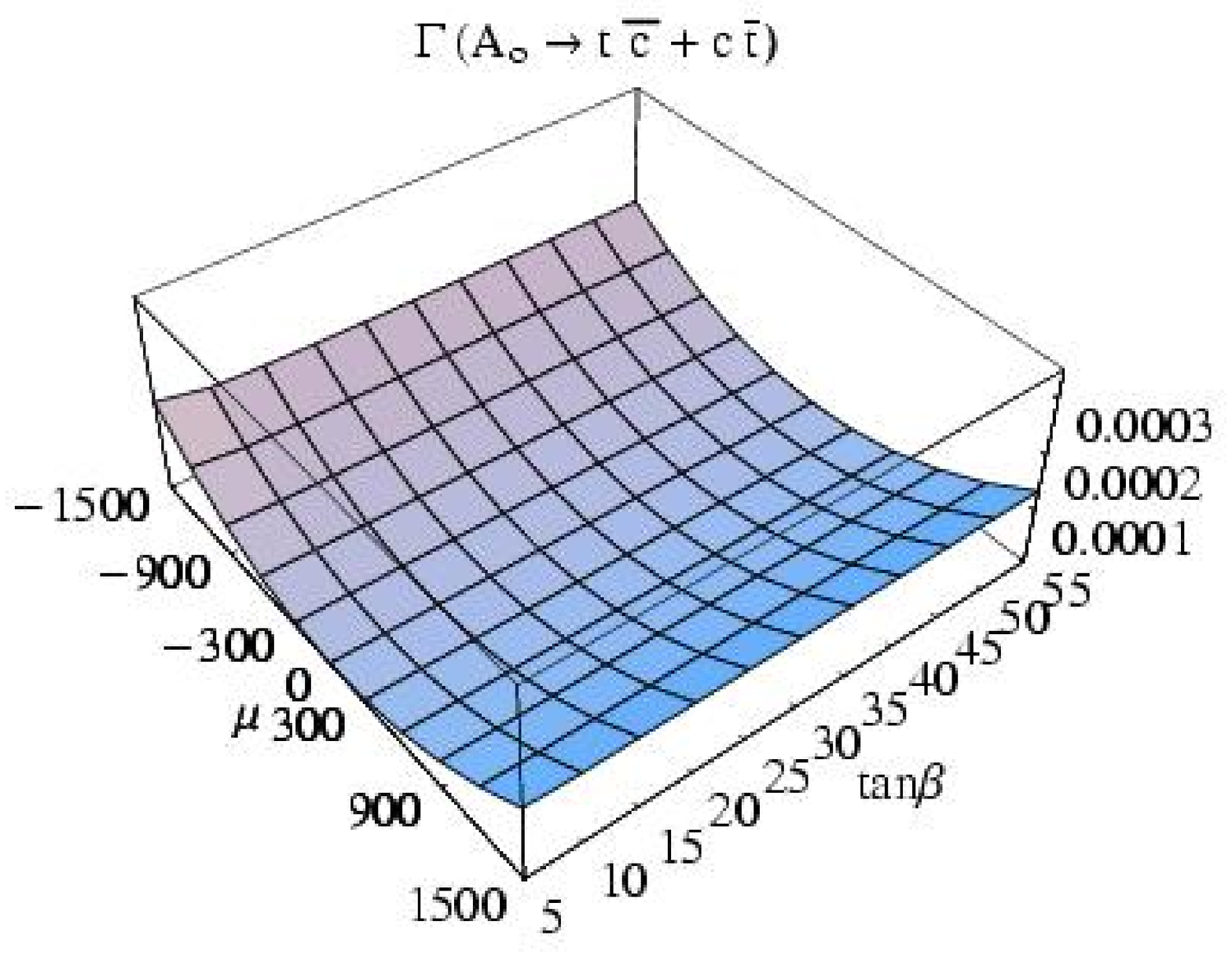,height=2.5in}\\
\vspace*{0.75cm}
\vspace{-1.35cm}
\hspace{-1.0cm}
\epsfig{file=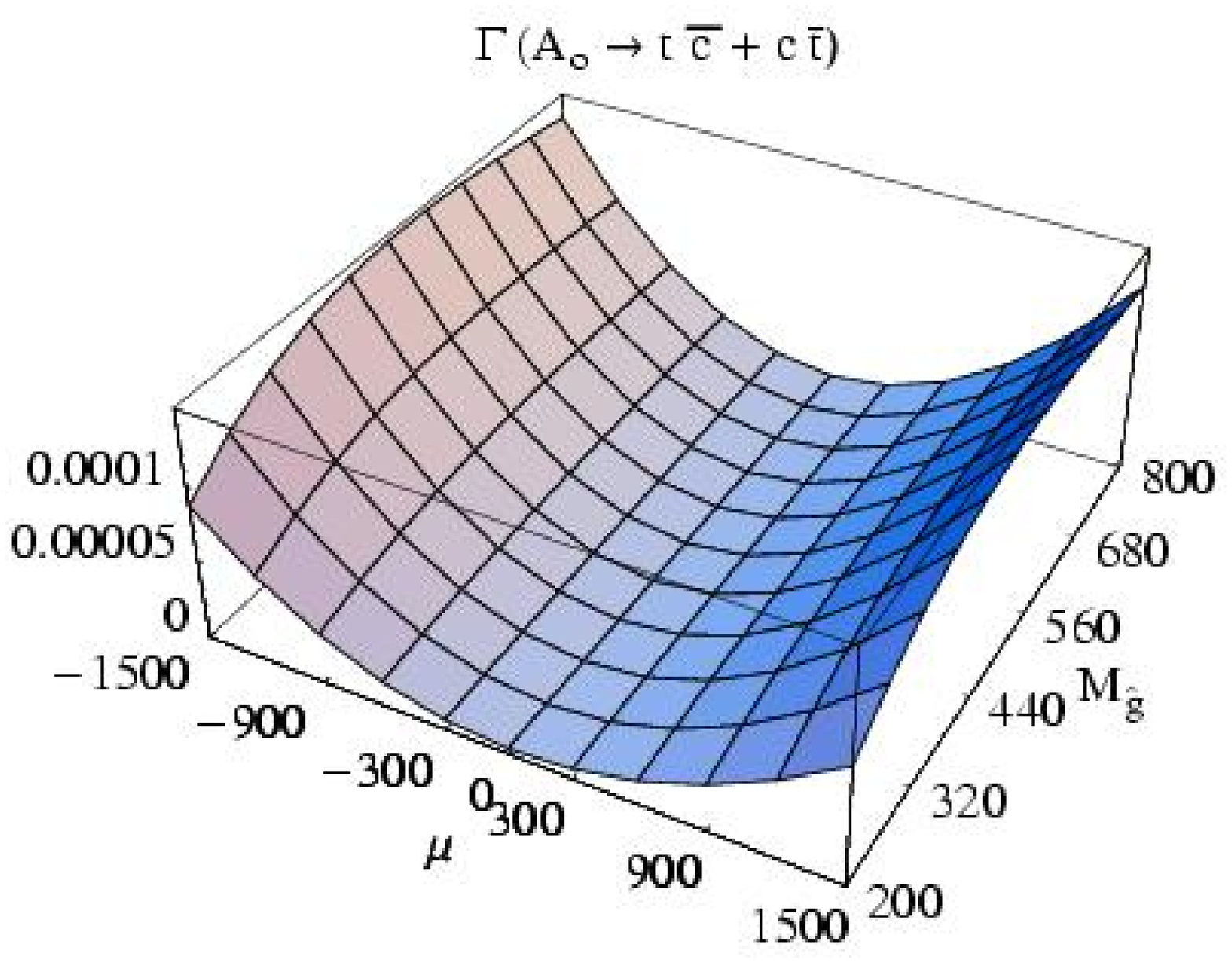,height=2.5in}
\epsfig{file=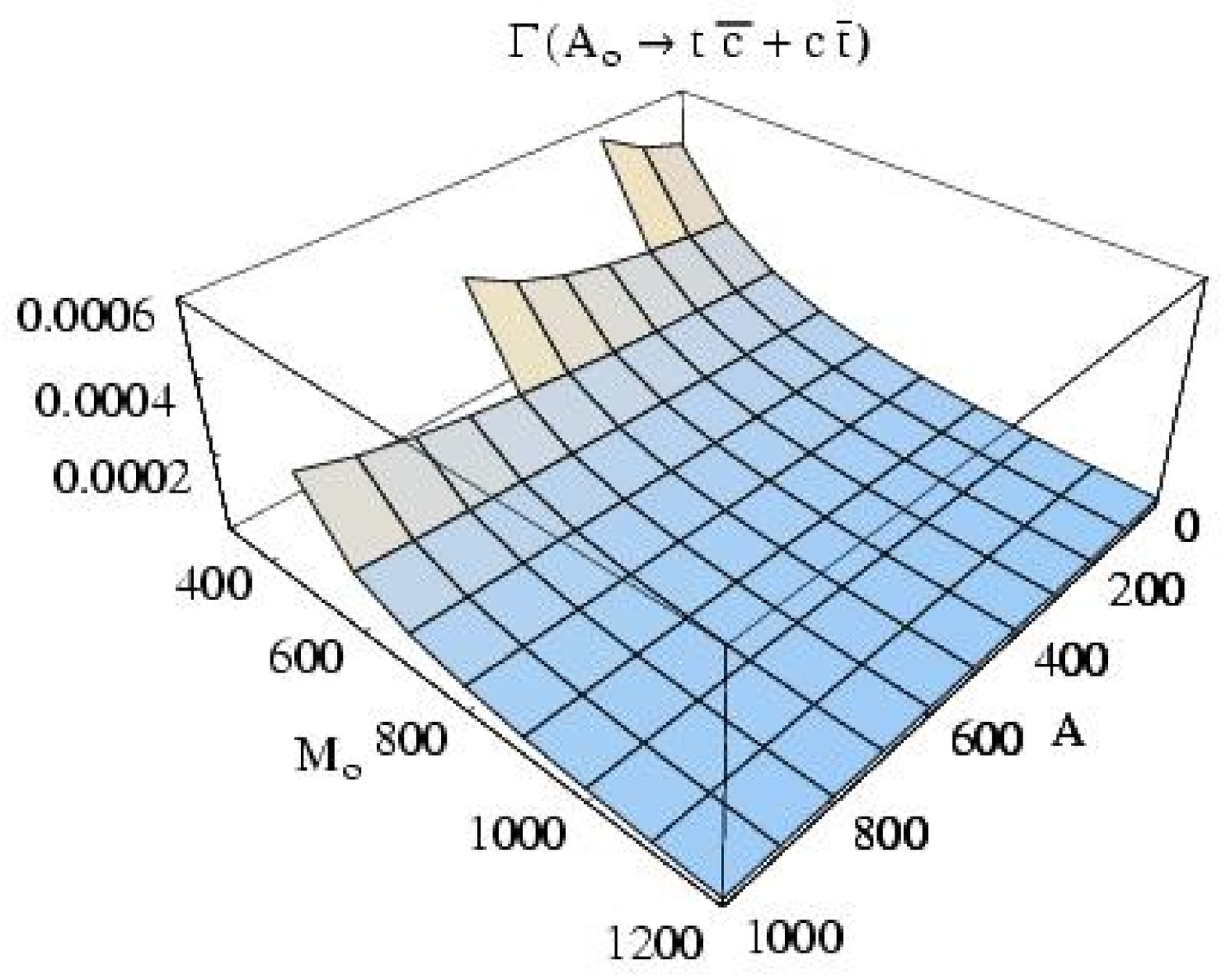,height=2.5in}\vspace{0.4em}
\caption{Same as in fig.\ref{htc_2} but for $\Gamma (A_o \to t \bar c + c \bar t)$. }
\label{htc_3}
\end{figure}
\end{center}
\vspace*{-1.05cm}

The figs.\ref{hbs_1} through \ref{htc_3} display the numerical results for the 
$\Gamma (H_a \to b \bar s+ s \bar b)$ and
$\Gamma (H_a \to t \bar c+ c \bar t)$ partial widths as a function of the six
previous 
MSSM parameters, $m_A$, $\tan \beta$, $\mu$, $M_{\tilde g}$, $M_0$ and $A$, and
for the specific value $\lambda=0.5$. In these figures, the MSSM parameters 
have been grouped in
pairs  
in order to illustrate simultaneously the behaviour of the FCHD widths 
with both parameters. These pairs of parameters have been chosen to be: 
($\tan \beta$, $m_A$), ($\mu$, $\tan \beta$), 
($\mu$, $M_{\tilde g}$) and ($M_o$, $A$), and correspond to
the plots {\bf (a)} (upper left panel), 
{\bf (b)} (upper right panel), 
{\bf (c)} (lower left panel) 
and {\bf (d)} (lower right panel), 
respectively, in these
figures. 
We have made these four plots for each neutral Higgs boson. 
figs.\ref{hbs_1},~\ref{hbs_2} and \ref{hbs_3} correspond to 
$\Gamma (h_o \to b \bar s+ s \bar b),\, \Gamma (H_o \to b \bar s+ s \bar b)$ 
and $\Gamma (A_o \to b \bar s+ s \bar b)$, respectively, and figs.\ref{htc_2} 
and \ref{htc_3} correspond to $\Gamma (H_o \to t \bar c+ c \bar t)$ and 
$\Gamma (A_o \to t \bar c+ c \bar t)$, respectively. Notice again that the 
lightest Higgs cannot decay to $t \bar c$ nor $c \bar t$, due to phase space.
 
In the following we
 discuss separately the $\Gamma (H_a \to b \bar s+ s \bar b)$ and 
 $\Gamma (H_a \to t \bar c+ c \bar t)$ cases due to their different behaviour 
 with some of the MSSM parameters.

First, we will focus on $\Gamma ( H_a \to b \bar s + s \bar b)$ 
(see figs.\ref{hbs_1}, \ref{hbs_2} and \ref{hbs_3}, 
for the $h_o$, $H_o$ and $A_o$ respectively). Let us start with the first plot,
{\bf (a)}, of these figures
where we study the combined behaviour with $\tan \beta$ and $m_A$.   
The first clear behaviour of these decay widths is the fast growing 
with $\tan \beta$. 
Notice that it is the same behaviour for the three neutral Higgs bosons, 
despite the fact that the growing with $\tan \beta$ is better seen in the 
case of the $H_o$ and $A_o$ than in the lightest Higgs one, due only to 
the different scales shown. It is important to recall that there are some
 values of the widths that are not shown in the figures since they correspond 
 to non allowed values for the squark masses, that is 
 $M_{\tilde q} < 150 \, GeV$.
  In particular, in this first plot we can see that for the values of the 
  parameters specified in the figure caption, values of about 
  $\tan \beta >40$ are not allowed in order to keep the squarks masses in 
  the allowed region.
Keeping this in mind and the growing behaviour with $\tan \beta$, 
we can conclude that the searched parameter space region that maximizes 
the FC effect is localized at large $\tan \beta$ values, with the largest 
possible values being constrained to be below some maximum, which depends 
on the particular values of the other parameters.

We next study the behaviour with $m_A$. We see again from plot {\bf (a)} of 
 each figure that it is clearly different, depending on the particular Higgs 
 decay we are studying. The decay widths $\Gamma (H_o \to b \bar s + s \bar b)$ 
 and $\Gamma (A_o \to b \bar s + s \bar b)$ clearly grow with $m_A$ due to 
 the obvious  phase space effects, that is, as $m_A$ increases, 
 the corresponding Higgs mass increases too. In contrast, the 
 $\Gamma (h_o \to b \bar s + s \bar b)$ width shows a less obvious behaviour 
 with $m_A$. As the lightest Higgs mass starts growing with $m_A$ and then it
  stabilizes, its decay width first grows with $m_A$, then it reaches a 
  maximum value and finally decreases with $m_A$, and we appreciate the 
  setting of the decoupling behaviour for large $m_A$ values that will be
   better explained in section~\ref{chap.decoupling}.

The phenomenologically interesting $m_A$ values would be those that can 
allow the next planned colliders to detect and study all the three neutral 
Higgs bosons. For definiteness, in our numerical analysis whenever we have 
to fix $m_A$, we have chosen $m_A = 250 \, GeV$, which for the interesting 
large $\tan \beta$ region, gives the three boson masses $m_{h_o}$, $m_{H_o}$, 
and $m_{A}$ being accessible to the LHC. On the other hand, regarding 
particularly the lightest Higgs, as it is the one that has more possibilities 
of being detected in the next years, it would be also interesting to further 
study the region of lower $m_A$ values but we do not explore this here.  
 
Now we focus on the behaviour with $\mu$. We can clearly see in 
figs.\ref{hbs_1}, \ref{hbs_2} and \ref{hbs_3}, {\bf (b)} and {\bf (c)}, 
that the widths are approximately symmetric under $\mu \to -\mu$ and that 
for moderate $|\mu| < 600 \, GeV$ they all grow with this parameter. We can 
appreciate this growing behaviour in all the flavour changing Higgs decays. 
The $H_o$ and $A_o$ decay widths grow with $|\mu|$ for all $\mu$ values,
 (figs.\ref{hbs_2} 
and \ref{hbs_3} {\bf (b)} and {\bf (c)}); but the lightest Higgs 
decay first grows with 
$|\mu|$ from $\mu = 0$, then, it reaches a maximum, decreases up to a minimum, 
and finally it continues growing (fig.\ref{hbs_1} {\bf (b)} and {\bf (c)}). 
We have found that this particular behaviour of the lightest Higgs boson 
is due to an accidental numerical cancellation among the contributions from 
the different diagrams to the form factors, $F_L^{bs}(h_o)$ and 
$F_R^{bs}(h_o)$, 
which takes place in this decay and not in the rest of decays under study.
 
 On the other hand, this behaviour with $\mu$, as can be seen in 
 figs.(\ref{hbs_1}, \ref{hbs_2} and \ref{hbs_3} {\bf (b)}), depends 
 also strongly on $\tan \beta$, because these two parameters appear 
 together in the squark mixings, and the behaviour of the widths with them 
 are correlated. For large $\tan \beta$ they grow faster with $\mu$, and for 
 small $\tan \beta$ the trilinear term starts to play a role too, 
 so this growing smoothes down. Due to this correlation, one has to be 
 careful in choosing independently both parameters as this could lead to 
 non allowed values of the squark masses, $M_{\tilde q} < 150 \, GeV$. 
 In particular, we can see again in plots {\bf (b)}, how for large values of 
 the $\mu$ parameter, as for instance $\mu = 1500 \, GeV$, only $\tan \beta $ 
 values below about 40 are allowed.

Following with $M_{\tilde g}$ (figs.\ref{hbs_1}, \ref{hbs_2} 
and \ref{hbs_3} {\bf (c)}), we find that the FC widths grow 
with $M_{\tilde g}$ up to a certain value $M_{\tilde g_o}$ and then they 
decrease very slowly (slow decoupling). Clearly, the value of $M_{\tilde g_o}$ 
depends 
on the rest of parameters. For the explored MSSM parameter space 
region, 
 we can see from these figures that it is in the range between about 
$200 \, GeV$ and $800 \, GeV$.  

On the other hand, the behaviour with $M_o$ is very clear 
too (figs.\ref{hbs_1}, \ref{hbs_2} and \ref{hbs_3} {\bf (d)}). 
All the decay widths decrease with this parameter, for large $M_o$, as it was expected. 
When $M_o$ grows, the masses of the squarks inside the loops grow as well, 
thus the probability of producing such loop-induced processes decreases as 
they are suppressed by the squark masses in the propagators.

To finish, we can see as well in figs. {\bf (d)} that the partial widths in 
the $d$-sector are nearly independent of the trilinear parameters $A$ for all 
the studied decays.
\begin{figure}[t]
\begin{center}
\vspace{-0.65cm}
\epsfig{file=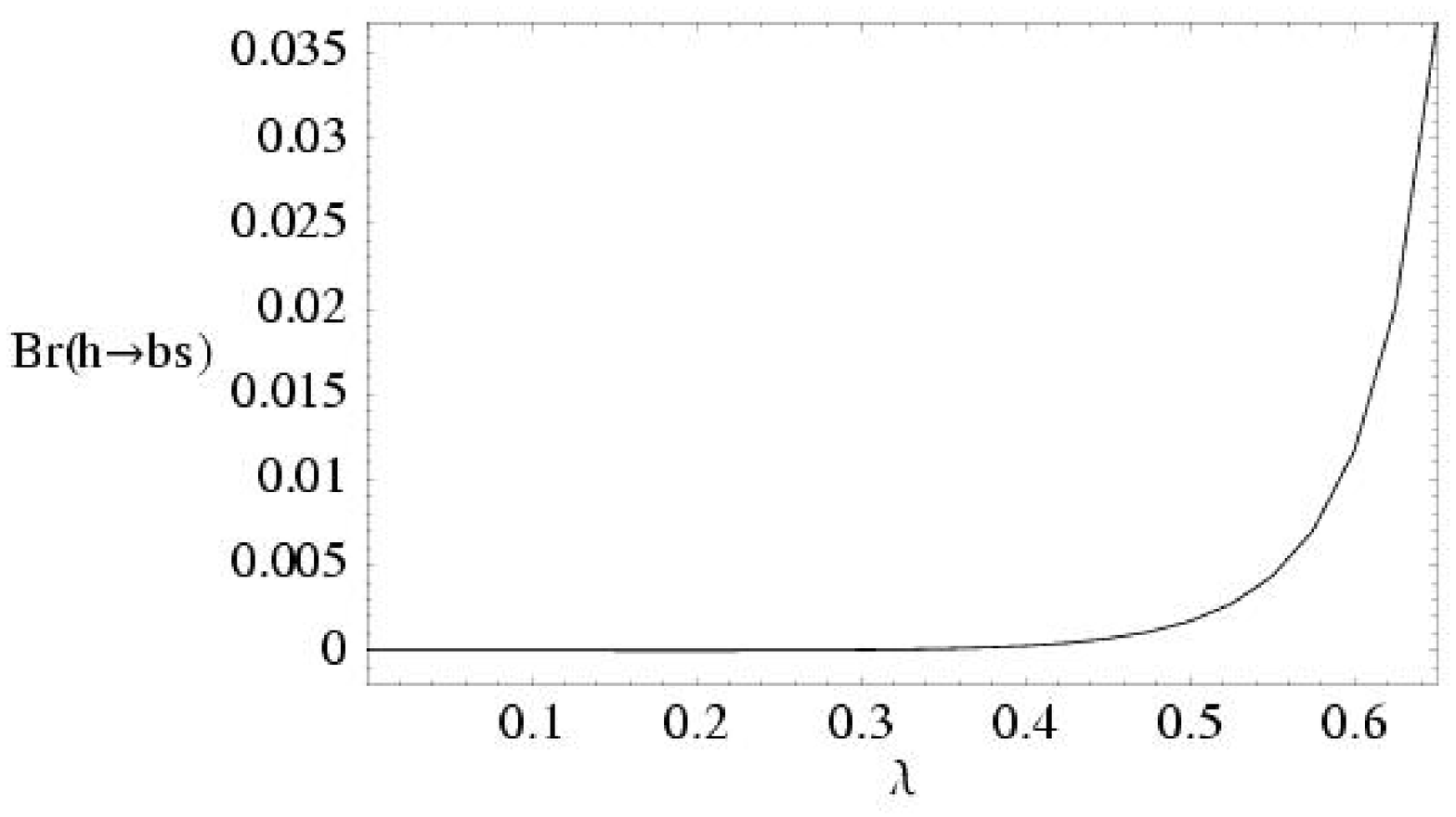,height=6.5cm,width=12cm} 
\epsfig{file=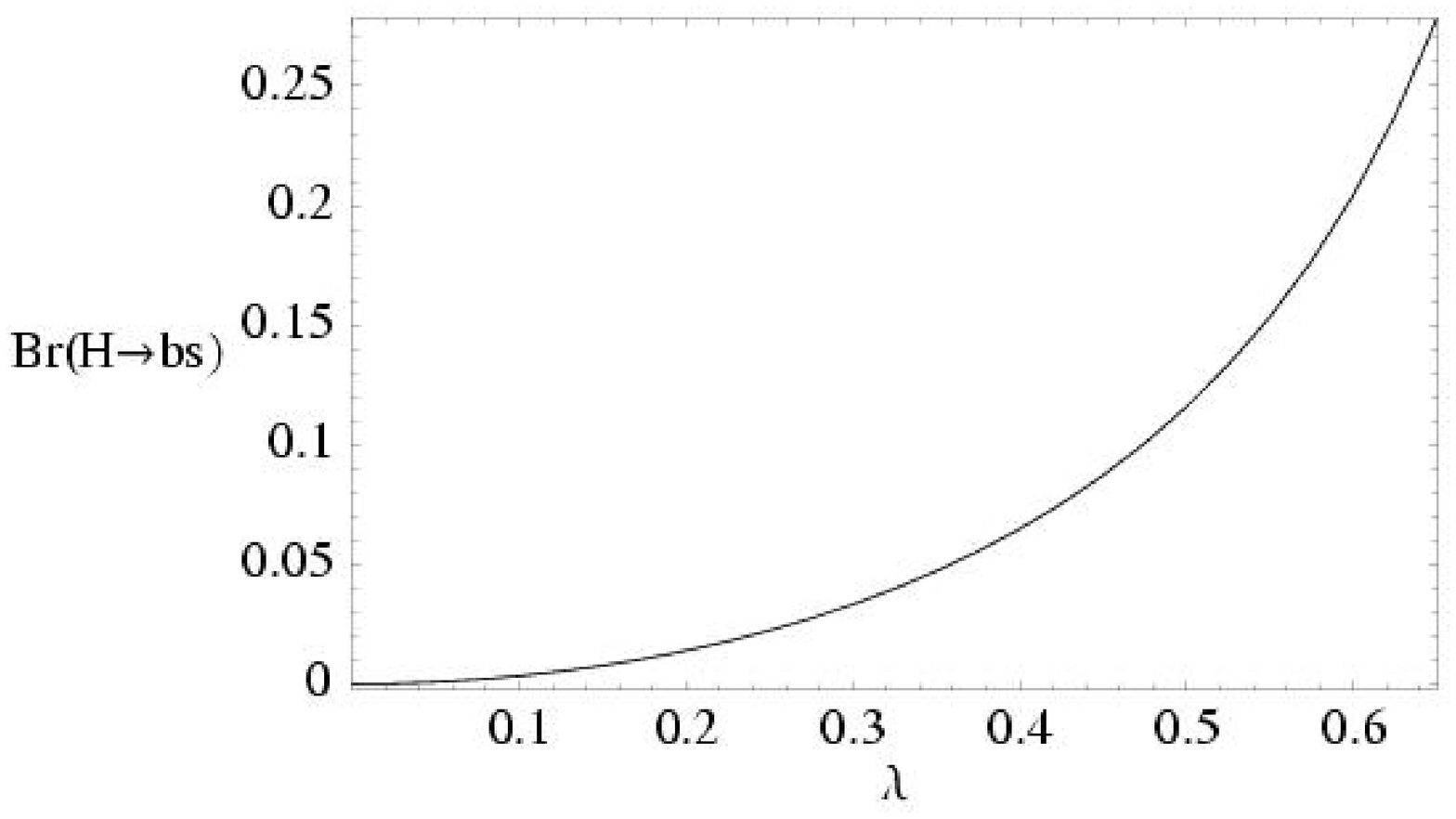,height=6.5cm,width=12cm}
\vspace{0.1in}
\caption{$Br( H_a \to b \bar s + s \bar b)$, $H_a =(h_o, H_o, A_o)$  as a function of 
$\lambda$ for the selected MSSM parameters. $Br(A_o \to b \bar s + s \bar b)$ is not plotted
 explicitly since in this
range of the parameter space is undistinguisable from the $H_o$ decay.
Higher values of $\lambda$ are not allowed 
as they give $M_{\tilde q} < 150 \, GeV$.}
\label{br_hbs_1}
\end{center}
\end{figure}
\begin{figure}[t]
\begin{center}
\vspace{-0.65cm}
\epsfig{file=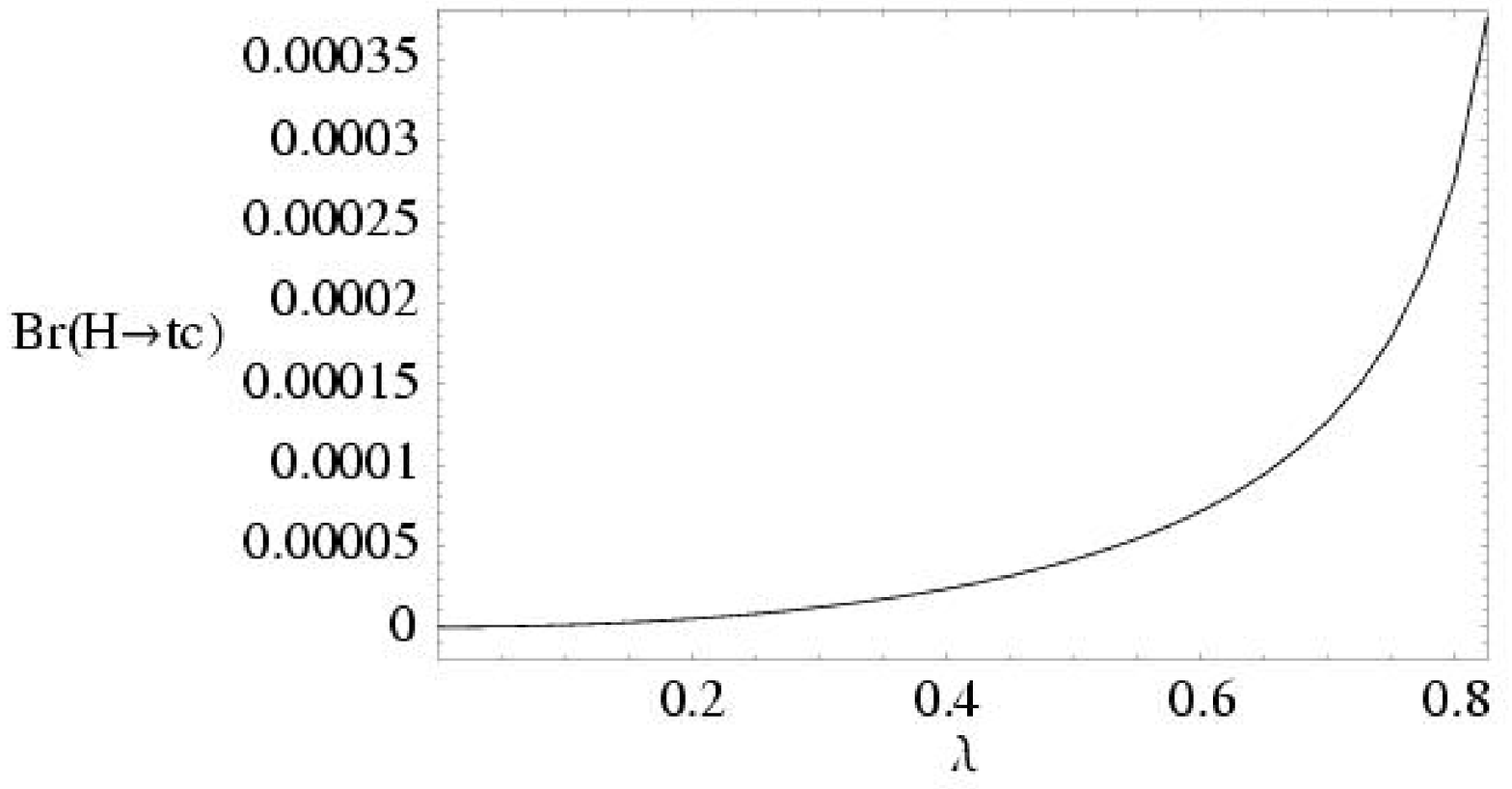,height=6.5cm,width=12cm} 
\epsfig{file=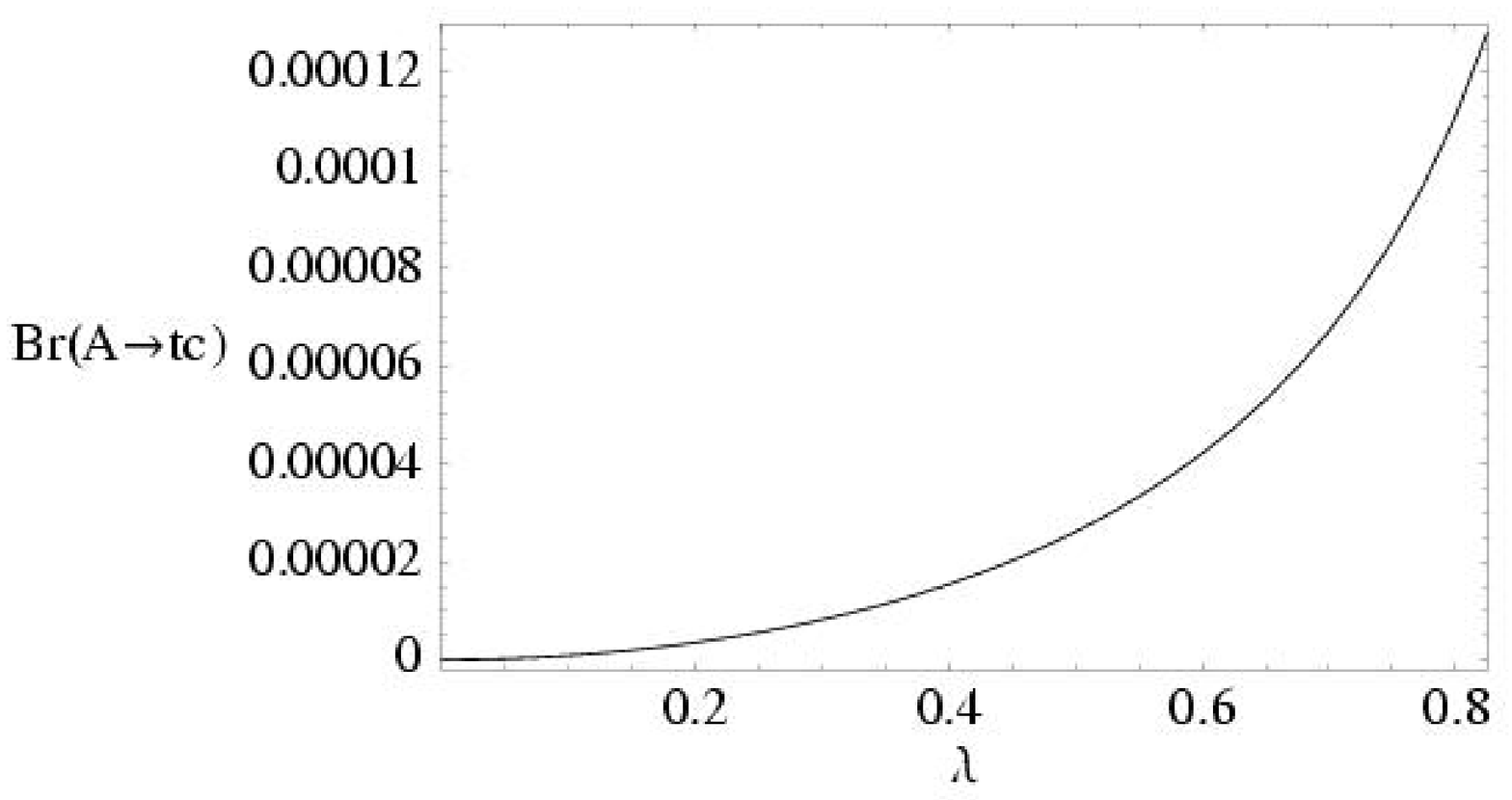,height=6.5cm,width=12cm}
\vspace{0.1in}
\caption{$Br( H_a \to t \bar c + c \bar t)$, $H_a = (H_o, A_o)$ as a function of $\lambda$ for
the selected MSSM parameters. Higher values of $\lambda$ are not allowed as they give $M_{\tilde
q} < 150 \, GeV$.}
\label{br_htc_2}
\end{center}
\end{figure}

In the following we will discuss the decay widths 
$\Gamma ( H_a \to t \bar c + c \bar t)$ (figs.\ref{htc_2} and \ref{htc_3}), 
where, as we have already explained, we will only study the behaviour of $H_o$ 
and $A_o$ decays, as the lightest Higgs cannot decay to $t \bar c$ or 
$c \bar t$. 

Again, the first plot {\bf (a)} of each figure represents the combined 
behaviour of the decay widths with $\tan \beta$ and $m_A$. One of the 
fundamental differences among these decays of the $u$-sector and the ones 
studied before of the $d$-sector is that now there is a very mild dependence 
with $\tan \beta$ and the widths decrease softly with $\tan \beta$.
Therefore, this time, the value that maximizes our effect is the smallest 
$\tan \beta$ value, but as we have said, the numerical effect of this parameter 
is not of crucial importance in these decays.

Regarding the behaviour with $m_A$, as can be seen in figs.\ref{htc_2} 
and \ref{htc_3} {\bf (a)}, it does not vary much with respect to the previous 
ones. The decay widths of the $H_o$ and $A_o$ again grow with $m_A$ due to 
the phase space as explained above, thus the same motivations to select a 
particular $m_A$ value will apply here. 

The partial widths again grow with $\mu$ and are approximately symmetric 
under $\mu \to -\mu$, but they show less symmetry than in the decays 
of the $d$-quark sector studied earlier. The squark mixing term now goes 
with $m_{t,c}(A_{t,c} - \mu \cot \beta)$ so the trilinear parameter $A$ becomes more 
important for $\tan \beta > 1$. Looking at the plots \ref{htc_2} 
and \ref{htc_3} {\bf (b), (c)} we can see that the maximum values are 
reached for negative values of $\mu$. We see as well that for large values 
of $\mu$ we can better appreciate the $\tan \beta$ dependence. 

Looking at the behaviour of the partial widths in the $u$-quark sector with 
the gluino mass, $M_{\tilde g}$, in figs.\ref{htc_2} 
and \ref{htc_3} {\bf (c)}, the widths again grow with $M_{\tilde g}$ until 
they reach a maximum at a certain value $M_{\tilde g_o}$, and then they start 
a slow decreasing (slow decoupling), as in the case of the $d$-quark sector 
studied above. Again, the value of $M_{\tilde g_o}$ also depends on the rest 
of parameters, and the mentioned behaviour is more evident for large values 
of the parameter $\mu$.

The behaviour with $M_o$ is again the same as before, decreasing with it as 
it was expected, but now, the trilinear parameter $A$ acquires more importance 
than before, growing the flavour changing widths as we increase the value of 
$A$. Again here, it is worth mentioning that there are some values not being 
plotted that correspond to non allowed values for the squark masses. 
In particular, for small values of $M_o$, only small values of $A$ are allowed.
 As a particular example, looking at figs.\ref{htc_2} 
 and \ref{htc_3} {\bf (d)} we can see that for 
 $M_o = 400 \, GeV$, values of the trilinear parameter $A$ larger than 
 about $600 \, GeV$ would not be allowed. 

Finally,  
we study the behaviour of the corresponding FCHD branching ratios with $\lambda$, 
and estimate their size for $0 \le \lambda \le 1$ and for 
some selected points of the MSSM parameter space, which have been chosen 
belonging to the 
region where the FC effects we are looking for are maximized. 
This region can be extracted easily from our results in the previous figures. 
Thus, for the $H_a \to b \bar s , s \bar b$ decays we select the following
representative point: $m_A=250 \, GeV$, $\tan \beta =35$, $\mu=1500 \, GeV$, 
$M_{\tilde g}=300 \, GeV$, $M_0=600 \, GeV$ and $A=200 \, GeV$; and  for the 
$H_a \to t \bar c , c \bar t$ decays we choose: $m_A=250 \, GeV$, 
$\tan \beta =10$, $\mu=-1500 \, GeV$, 
$M_{\tilde g}=500 \, GeV$, $M_0=600 \, GeV$ and $A=800 \, GeV$. Regarding 
the computation of the total Higgs boson widths, these    
have been evaluated with the HDECAY program~\cite{hdecay}.

For the $H_a \to b \bar s , s \bar b$ decays, as can be seen in 
fig.\ref{br_hbs_1}, the branching ratios grow with $\lambda$, being 
exactly zero for $\lambda=0$, as expected, and can reach quite sizeable 
values, even for moderate $\lambda$.
For instance, for $\lambda \approx 0.6$ the branching ratio for the $h_o$ is
about $0.015$, and for the $A_o$ and $H_o$ is about $0.25$. Of course, larger 
$\lambda$ values would lead to larger rates, but these are not allowed here 
due to the  $M_{\tilde q}>150 \, GeV$ restriction.  

For the $H_a \to t \bar c, \, c \bar t$ decays, as can be seen in 
fig.\ref{br_htc_2}, the branching ratios grow again with $\lambda$, 
being exactly zero for $\lambda=0$. Again, only $\lambda$ values leading to   
$M_{\tilde q}>150 \, GeV$ are shown.
The $H_o$ branching ratio reaches its maximum value, of about 0.00035 
for $\lambda \approx 0.8$, while the pseudoscalar branching ratio 
reaches 0.0001 for this same value of $\lambda$.  

In summary, the  FCHD branching ratios that we have found in this section 
are quite sizable, in fact, many orders of magnitude larger that the 
corresponding SM rates (for instance, in the $u$-sector, and for 
$m_{H_{SM}}= 200 \, GeV$ we estimate $B(H_{SM}\to t \bar c + c \bar t) \sim 10^{-13})$ 	 
and will produce an interesting amount of rare events at the planned next 
generation colliders. We do not study here this in more detail and 
postpone this interesting analysis for a future work.

%%%%%%%%%%%%%%%%%%%%%%%%%%%%%%%%%%%%%%%%%%%%%%%%%%%%%%%%%%%%%%
\section{Non-Decoupling Behaviour of Heavy Squarks and Gluinos in Flavour 
Changing Higgs Decays}
\label{chap.decoupling}

In the previous sections we have studied the FC effects of 
squarks and gluinos via SUSY-QCD radiative corrections in the decays of 
neutral Higgs particles within the MSSM. We have performed 
analytical and numerical calculations, and estimated the size of these   
effects as a function of the various MSSM parameters  and $\lambda$.  
In this section, we study the behaviour of these FCHD processes in the most 
pessimistic scenario where a heavy supersymmetric spectrum is considered, 
and we will show that these FC effects remain sizable even for squark and 
gluino masses as large as $\mathcal{O} (1 TeV)$. We will show that the reason 
for these corrections remaining so large is that at asymptotically large 
SUSY particle masses they manifest a non-decoupling behaviour. That is, the 
one loop SUSY-QCD radiative corrections to the neutral MSSM Higgs boson 
decays remain non-vanishing at asymptotically large squark and gluino masses, 
and it results in a finite and non-negligible contribution which in principle 
could provide an indirect signal of SUSY-QCD at low energies, even in this 
pessimistic scenario.

The decoupling of all SUSY particles would imply that the prediction in the
MSSM for all 
the observables involving non-SUSY particles in the external legs, such as 
the partial decay widths under study here, should tend, in the limit of 
large SUSY masses, to their corresponding 
values in the non-SUSY two Higgs doublet model, called 2HDMII in the
literature~\cite{2HDM}. Formally, and following the lines of the 
{\it Appelquist-Carazzone Theorem}~\cite{Apple}, the decoupling 
would occur
if the contributions of the SUSY particles to low-energy processes either 
fall 
as inverse powers of the SUSY mass parameters or can be absorbed into 
redefinitions of the couplings and parameters of the low-energy theory. 
In the present case, the decoupling of SUSY particles in the FCHD would imply,
in particular, 
that the  SUSY-QCD induced radiative corrections should tend to zero 
in the asymptotic 
large SUSY mass limit.
This theorem has been proved to be valid for theories 
with an exact gauge symmetry, however, it does not apply to 
theories with spontaneously broken gauge symmetries, nor with chiral fermions,
as it is clearly the case of the MSSM. Furthermore,  
in order to have decoupling, the dimensionless couplings should not grow 
with the heavy masses. Otherwise, the mass suppression induced by the 
heavy-particle propagators can be compensated by the mass enhancement 
provided by the interaction vertices, with an overall non-vanishing effect, 
which is exactly what happens in some Higgs boson decays to fermions. 
This has been proved to happen in some flavour preserving MSSM neutral Higgs 
boson decays in a series of previous works~\cite{HaberTemes,ourHtb,MJsitges,dobado,Ana,siannah}
and also in $H^+$ production at hadronic colliders~\cite{decinprod}. 
Here we will show that the non-decoupling effects also appear in the 
FCHD. These non-decoupling effects can have interesting phenomenological 
consequences as they imply that low-energy experiments can be sensitive 
to large mass scales, which cannot be kinematically accessed by direct searches. 
     
In order to show analitically the non-decoupling behaviour of squarks 
and gluinos in the FCHD, we perform a systematic expansion of the form 
factors involved, and hence
of the partial widths, in inverse powers of the heavy SUSY masses and look for
the first term in this expansion. To define our expansion parameter, 
we consider here the simplest hypothesis where all the soft-SUSY-breaking mass 
parameters and the $\mu$ parameter are all of the same order 
(collectively denoted by $M_{S}$) and much heavier than the electroweak 
scale, $M_{EW}$, in such a way
that the differences between these mass parameters are considered to be of 
order $M_{EW}$. That is,
\begin{equation}
M_{S} \simeq M_{o} \simeq M_{\tilde g} \simeq \mu \simeq A  
\gg M_{EW}
\label{eq.largeSUSYlimit}
\end{equation}
where again $M_o$ denotes the common soft breaking squark mass parameter, 
$M_o = M_{\tilde Q} = M_{\tilde U} = M_{\tilde D}$, and  
$M_{\tilde Q}$, $M_{\tilde U}$ and $M_{\tilde D}$ are chosen here, for
simplicity, to be the same for the two 
generations involved, and $A$ is the common trilinear parameter. 
Notice that, with $M_{S} \sim \mathcal{O}(1 TeV)$, this
choice will lead to a plausible situation where all the SUSY particles 
in the SUSY-QCD sector are much heavier
than their SM partners.\footnote{ Notice that the condition $\mu \sim M_{S}$ 
is not necessary to
get the SUSY-QCD particles heavy but is needed in order to get all 
the charginos and neutralinos heavy in the SUSY-electroweak sector.
The condition of large trilinear couplings $A_q \sim M_{S}$ is not 
necessary to get large SUSY masses.
However it is a plausible  choice from the theoretical perspective if one 
assumes that all the soft breaking
parameters have a common origin.}

In the following, we perform the expansions of the SUSY-QCD contributions to 
the form factors, whose exact analytical results where presented in the previous
section, in inverse
powers of $M_S$ and keep just the leading contribution of this expansion 
by considering that all the remaining involved mass scales 
$m_{H_o},m_{A},m_{h_o},m_Z,m_W$ and $m_q$  are of order $M_{EW}$. To this end,
we use the results of the expansions of the one-loop functions and the rotation
matrices that are given in Appendix B.   

For the $H_a \to b\bar s, \, s\bar b$ decays, with $H_a=h_o,H_o,A_o$, we find 
the following results for the leading terms 
in the expansions of the form factors,  
\begin{eqnarray}
F_L^{bs} (h_o) &=& \frac{\alpha_s}{6 \pi} \frac{m_b \sin\alpha}{2 m_W \cos\beta} 
(\tan\beta + \cot\alpha) \frac{\mu M_{\tilde g}}{M_0^2} F(\lambda) \nn \\
F_R^{bs} (h_o) &=& \frac{\alpha_s}{6 \pi} \frac{m_s \sin\alpha}{2 m_W \cos\beta} 
(\tan\beta + \cot\alpha) \frac{\mu M_{\tilde g}}{M_0^2} F(\lambda) \nn \\
F_L^{bs} (H_o) &=& - \frac{\alpha_s}{6 \pi} \frac{m_b \cos\alpha}{2 m_W \cos\beta} 
(\tan\beta - \tan\alpha) \frac{\mu M_{\tilde g}}{M_0^2} F(\lambda) \nn \\
F_R^{bs} (H_o) &=& - \frac{\alpha_s}{6 \pi} \frac{m_s \cos\alpha}{2 m_W \cos\beta} 
(\tan\beta - \tan\alpha) \frac{\mu M_{\tilde g}}{M_0^2} F(\lambda) \nn \\
F_L^{bs} (A_o) &=& -i \frac{\alpha_s}{6 \pi} \frac{ m_b}{2 m_W} \tan\beta 
(\tan\beta + \cot\beta) \frac{\mu M_{\tilde g}}{M_0^2} F(\lambda) \nn \\
F_R^{bs} (A_o) &=& i \frac{\alpha_s}{6 \pi} \frac{ m_s}{2 m_W} \tan\beta 
(\tan\beta + \cot\beta) \frac{\mu M_{\tilde g}}{M_0^2} F(\lambda) 
\label{eq.hbsLLonescale} 
\end{eqnarray}

Similarly, for the $H_a \to t\bar c, \, c\bar t$ decays, with $H_a=H_o,A_o$, 
we find the following results for the leading terms 
in the expansions of the form factors,  
\begin{eqnarray}
F_L^{tc} (H_o) &=& - \frac{\alpha_s}{6 \pi} \frac{m_t \sin\alpha}{2 m_W \sin\beta} 
(\cot\beta - \cot\alpha) \frac{\mu M_{\tilde g}}{M_0^2} F(\lambda) \nn \\
F_R^{tc} (H_o) &=& - \frac{\alpha_s}{6 \pi} \frac{m_c \sin\alpha}{2 m_W \sin\beta} 
(\cot\beta - \cot\alpha) \frac{\mu M_{\tilde g}}{M_0^2} F(\lambda) \nn \\
F_L^{tc} (A_o) &=& -i \frac{\alpha_s}{6 \pi} \frac{ m_t}{2 m_W} \cot\beta 
(\tan\beta + \cot\beta) \frac{\mu M_{\tilde g}}{M_0^2} F(\lambda) \nn \\
F_R^{tc} (A_o) &=& i \frac{\alpha_s}{6 \pi} \frac{ m_c}{2 m_W} \cot\beta 
(\tan\beta + \cot\beta) \frac{\mu M_{\tilde g}}{M_0^2} F(\lambda) 
\label{eq.htcLLonescale} 
\end{eqnarray}

\begin{figure}[t]
\begin{center}
\hspace{-0.5cm}
\epsfig{file=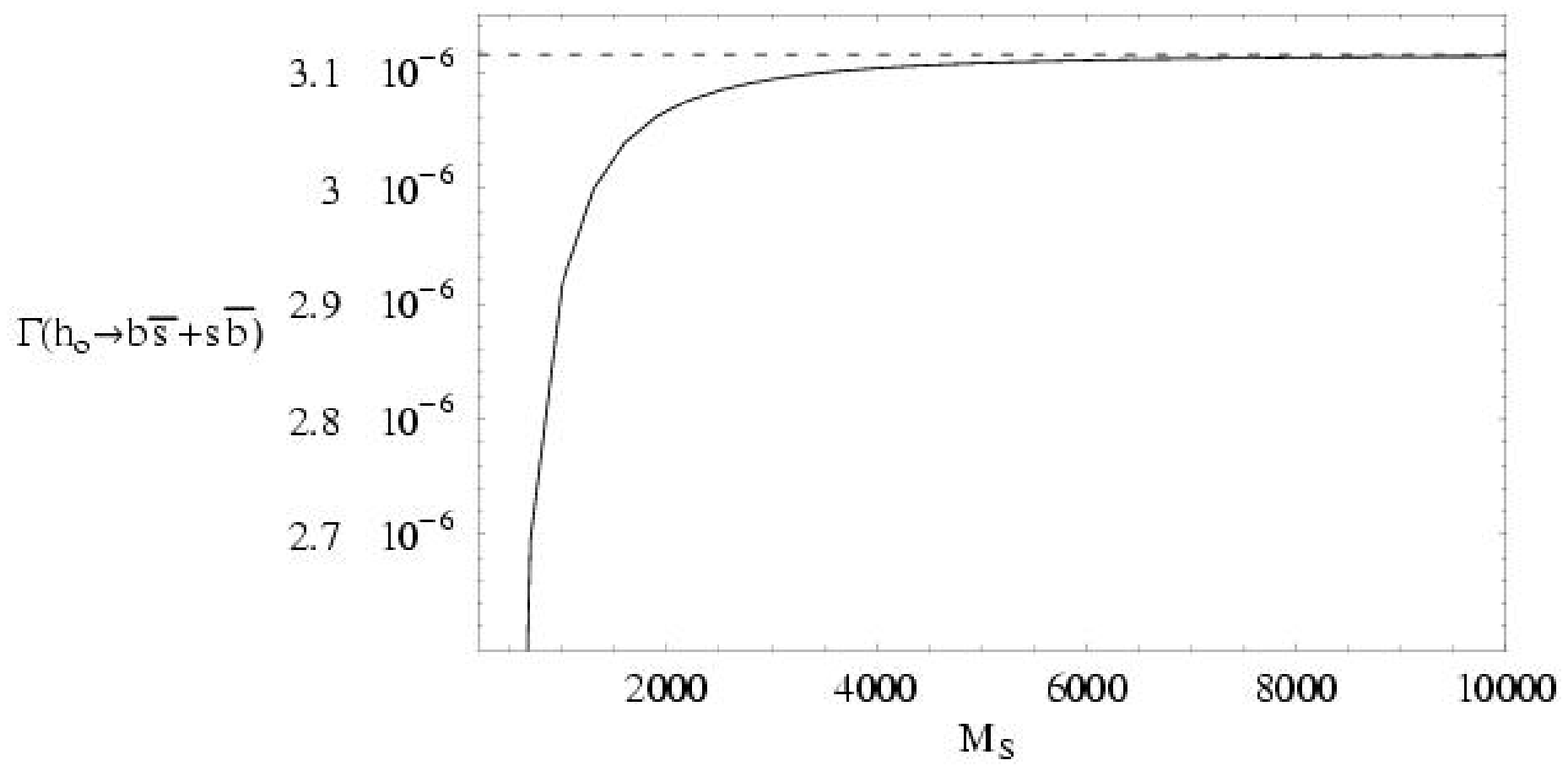,height=6cm,width=13cm}
\epsfig{file=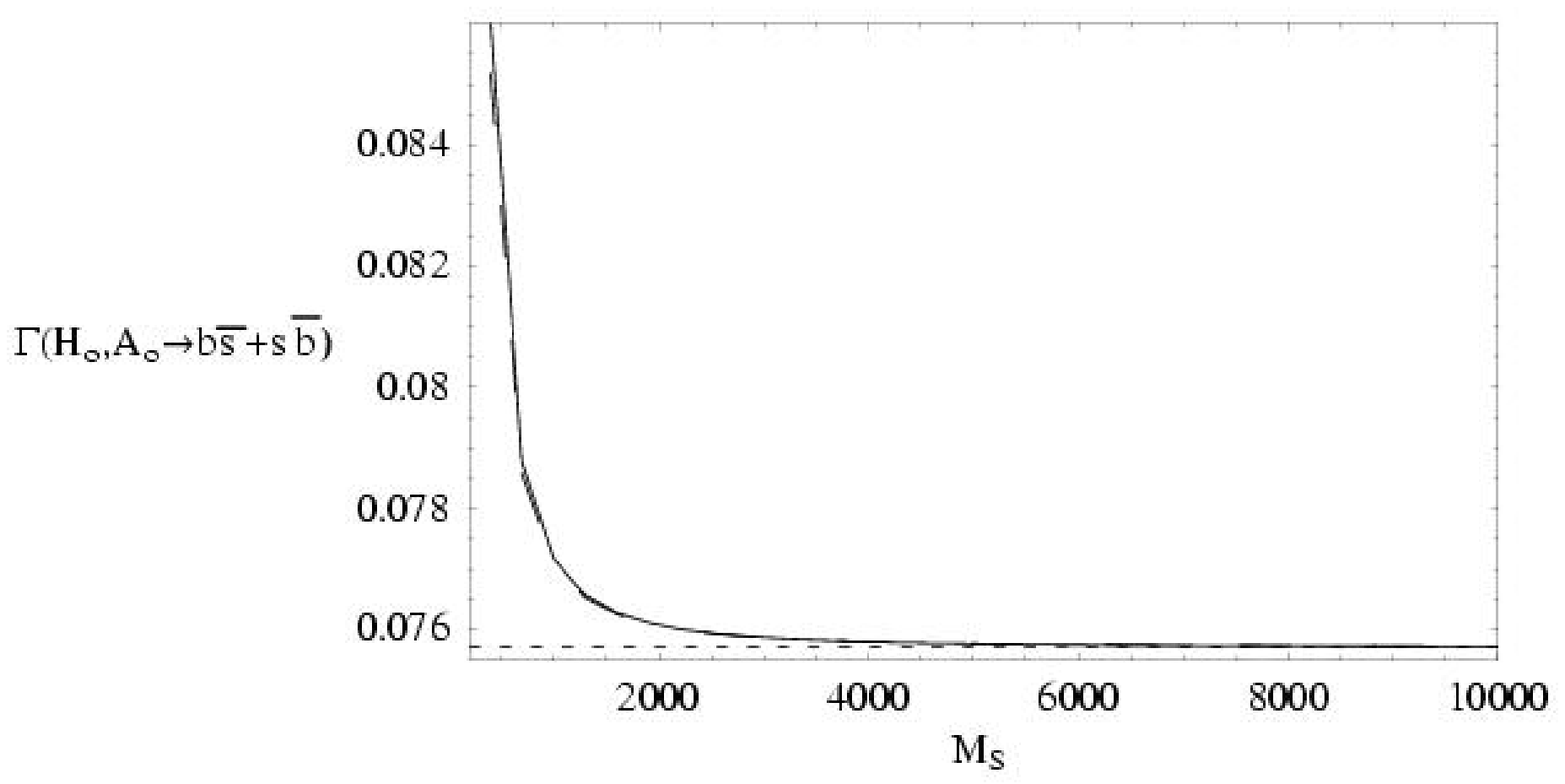,height=6cm,width=13cm}
\caption{\it Non-decoupling behaviour of $\Gamma(H_a \to b\bar s + s\bar b)$ in $GeV$ with 
$M_0=\mu=A=M_{\tilde g}=M_S$, 
for $H_a = h_o$ (top panel) and $H_a =H_o, A_o$ (bottom panel) 
and for $\tan\beta=35$, $\lambda = 0.5$, $m_{A}=m_{H_o}=250 \, GeV$ and $m_{h_o}=135 \, GeV$.
 Exact one-loop results in solid lines for the $h_o$ and $H_o$, long-dashed lines for the
  $A_o$. The expansions
given in eq.(\ref{eq.hbsLLonescale}) (short-dashed) are plotted for comparison.}
\label{fig.hbsLLonescale}
\end{center}
\end{figure}
\begin{figure}[h]
\begin{center}
\epsfig{file=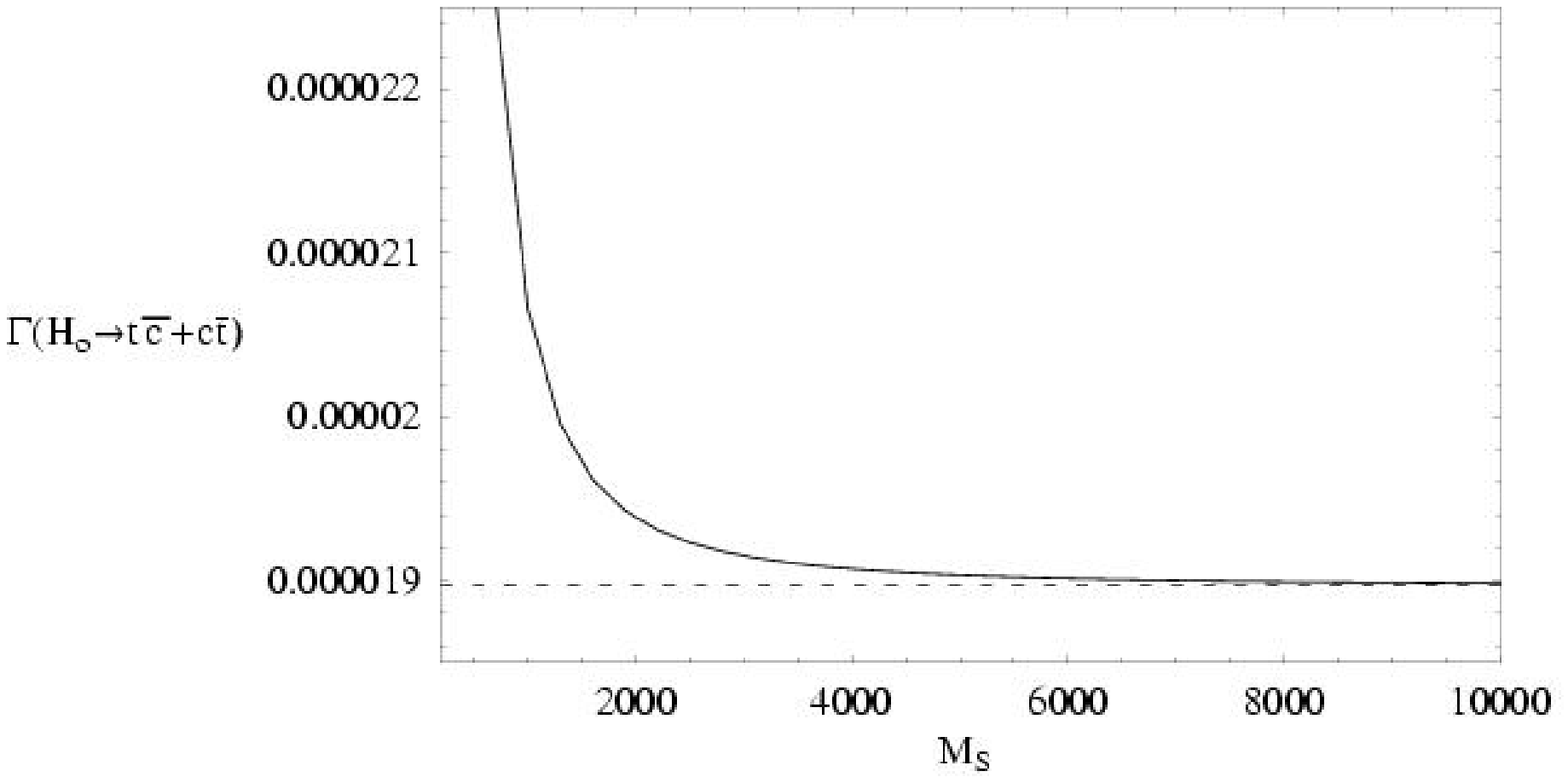,height=6cm,width=13cm}
\caption{\it Non-decoupling behaviour of $\Gamma(H_o \to t\bar c
+c\bar t)$ in $GeV$ with
 $M_0=\mu=A=M_{\tilde g}=M_S$ 
and for $\tan\beta=35$, $\lambda = 0.5$ and $m_{H_o}=250 \, GeV$. 
Exact one-loop results 
(solid) and the expansions
given in eq.(\ref{eq.htcLLonescale}) (short-dashed lines) are plotted
for comparison. $\Gamma(A_o \to t\bar c
+c\bar t)$ is not plotted explicitly since in this range of the
parameter space is undistinguisable from the $H_o$ decay.}
\label{fig.htcLLonescale}
\end{center}
\end{figure}

Some comments are in order. 
First, we can see from eqs.(\ref{eq.hbsLLonescale}) and 
(\ref{eq.htcLLonescale}) that, taking all SUSY mass parameters arbitrarily 
large and of the same
order, the SUSY-QCD contributions to the FCHD partial widths lead to a 
non-zero value. That is, they do not decouple in
the large SUSY mass scenario. This can be seen clearly, for instance, 
in the simplest case of equal mass
scales, $\mu=M_{\tilde g}=M_0$, in which the dependence on the SUSY mass 
scale of the leading contributions disappear, and these provide a constant 
and non-vanishing value. 

In the above eqs.(\ref{eq.hbsLLonescale}) and (\ref{eq.htcLLonescale}) there
appear one new function that is defined as, 
\begin{equation}
F(\lambda) = \frac{2}{\lambda^2} [(\lambda +1)\ln(\lambda+1) + 
(\lambda -1)\ln(1-\lambda) - 2\lambda],
\end{equation}
where $\lambda$ is the parameter 
introduced in eqs.(\ref{eq.usquarkmass}) and (\ref{eq.dsquarkmass}), and it is
the only function carrying the information of the FC strength in our 
asymptotic results. Notice, that it gives a good approximation of the behaviour 
with $\lambda$ of the 
exact result for the FCHD branching ratios found in the previous section and
shown in figs.\ref{br_hbs_1} and \ref{br_htc_2}.
For small $\lambda$ values, $\lambda\ll 1$, it behaves as 
$F(\lambda)\simeq -2\lambda/3-\lambda^3/5$, and the first term, which is linear
in $\lambda$ provides the corresponding result if the mass insertion
approximation would have been used instead.

The previous results of eqs.(\ref{eq.hbsLLonescale}) and 
(\ref{eq.htcLLonescale}) are valid for all $m_A$ and $\tan \beta$ values and 
keep all the involved quark masses, $m_t$, $m_b$, $m_c$ and $m_s$, different
from zero.

These expressions are of 
$\mathcal{O}\left[(\frac{M_{EW}}{M_S})^0 \right]$ in the large SUSY mass 
expansion and have corrections coming from the next to leading terms that are 
of order $\left[(\frac{M_{EW}}{M_S})^n \right]$ with $n>0$. These latter are 
not shown explicitly since they vanish in the asymptotically large SUSY mass 
limit and therefore, they decouple. 

The results for the $F_L^{bs}$ form factors in 
eq.(\ref{eq.hbsLLonescale}) are in agreement with those in the literature
regarding Higgs-mediated FCNC effects, being radiatively induced from SUSY-QCD
particles, within the context of $B$ meson physics~\cite{toharia,kolda,newchanko,oldchanko}. These works use instead the effective Lagrangian approach, or,
equivalently, the zero external momentum approximation, are focused on the 
large $\tan \beta$ limit, and use the mass insertion approximation,
valid for small $\lambda \ll 1$ values, to describe 
the flavour mixing of the internal squark propagators. Our results are more
general in that they are for arbitrary $\tan \beta$ and $\lambda$ values, and
they recover the previous results when the limits $\tan \beta \gg 1$ and
$\lambda\ll 1$ are considered in eq.(\ref{eq.hbsLLonescale}). In addition,  
we include the predictions for the $F_R^{bs}$ form factors which are 
proportional to $m_s$ and are usually neglected. In eq.(\ref{eq.htcLLonescale}) we also provide 
predictions
for the form factors of the $u$-sector, $F_L^{tc}$ and $F_R^{tc}$, 
which are proportional to $m_t$ and $m_c$, respectively, and are obviously not relevant for the 
$B$-meson physics but can have interesting applications for future  
indirect SUSY searches at the top quark physics.  
     
It deserves special mention the convergence, 
in the $m_{A}>>m_Z$ limit~\cite{decoupling}, of our result for the $h_o$ form factors to the 
vanishing tree-level prediction of the SM rate. Actually, since $\cot
 \alpha \to - \tan \beta$ in this large $m_{A}$ limit, we can see this
 vanishing explicitly in $F_{L,R}^{bs}(h_o)$ of eq.(\ref{eq.hbsLLonescale}). 
        
Finally, in order to show numerically this non-decoupling behaviour, we plot
in fig.\ref{fig.hbsLLonescale} the exact one-loop results for the 
$H_a \to b\bar s, \, s\bar b$ partial widths and 
in fig.\ref{fig.htcLLonescale} for the $H_a \to t\bar c, \, c\bar t$ partial 
widths 
as functions of the large SUSY mass scale, $M_S$ (solid
lines for all except for $A_o \to b \bar s, \, s \bar b$ 
that is plotted by long-dashed lines). For comparison, we also plot the
approximate results from the leading
terms of our expansions in eqs.(\ref{eq.hbsLLonescale}) and
(\ref{eq.htcLLonescale}) (short-dashed lines). The values of $M_S$ have been taken 
here 
up 
to unrealistic very large values just to illustrate the convergence of the exact
and approximate results.  
 In all these plots, we choose 
 $M_0 = M_{\tilde g} = \mu = A=M_S$ and 
$\tan\beta=35$, $\lambda=0.5$. 
 We also fix $m_{h_o} = 135 \, GeV$, $m_{H_o} = 250 \, GeV$ and $m_{A} = 250 \, GeV$. 
 We can see in these figures 
that for large SUSY mass parameters, the  exact partial widths tend to a 
non-zero value and can be quite large even for
quite heavy SUSY particles. For example, 
$\Gamma (h_o \to b\bar s + s\bar b)$ (top panel in 
fig.\ref{fig.hbsLLonescale}) can be as large as $3 \times 10^{-6} \, GeV$ 
for $M_S \simeq 10 \, TeV$. The 
$H_o/A_o \to b\bar s, \, s\bar b$ cases are plotted together 
(bottom panel in fig.\ref{fig.hbsLLonescale}) as they are very similar, and can be as large as 
$7.6 \times 10^{-2} \, GeV$ for $M_S \simeq 10 \, TeV$.
The same behaviour can be seen for $\Gamma (H_o/A_o \to t\bar c +
c\bar t)$,
fig.\ref{fig.htcLLonescale}, 
but these are smaller than in their $bs$ decays, because
of the phase space suppression and the unfavourable value of large 
$\tan \beta$. 
For example for $M_S \simeq 10 \, TeV$ they are $1.9 \times 10^{-5} \, GeV$.

In summary, the non-decoupling behaviour found in this section explains 
the large values for the FCHD partial widths obtained
in the previous section for large values of the MSSM mass parameters. 
It is worth noticing that this non-decoupling property of the SUSY particles 
is associated to the fact that the mass suppression induced by the heavy 
sparticle propagator is being compensated because 
there is a Higgs-squark-squark coupling,
eqs.(\ref{ghsusu1}), (\ref{ghsusu2}), (\ref{ghsdsd1}) and (\ref{ghsdsd2}), 
with mass dimension that depends on $\mu$ and on $A$ and therefore grows
with $M_S$ in the large SUSY mass limit. On the other hand, from the 
numerical comparison between the exact and 
our approximate results of 
eqs.(\ref{eq.hbsLLonescale}) and (\ref{eq.htcLLonescale}), we
can conclude from fig.\ref{fig.hbsLLonescale} and 
fig.\ref{fig.htcLLonescale} that these large $M_S$ expansions, 
with just the leading term, are a good approximation for large enough SUSY 
mass parameters.  These asymptotic results for the FCHD form factors can have
interesting applications for future indirect SUSY searches at next generation
colliders, which together with the previously found non-decoupling effects in
flavour preserving Higgs decays~\cite{HaberTemes,ourHtb,MJsitges,dobado,Ana,siannah} and $H^+$ production
~\cite{decinprod}, could provide an indirect signal 
of SUSY, even
in the most pessimistic scenario of a heavy SUSY spectrum at the $TeV$ scale.  

%%%%%%%%%%%%%%%%%%%%%%%%%%%%%%%%%%%%%%%%%%%%%%%%%%%%%%%%%%%%%%%%%%%%%%%%%%%%%%
\section{ Summary and conclusions}
Our main goal in this work has been to study  the Flavour Changing Neutral 
Higgs Decays in the MSSM 
 as a possible indirect search for supersymmetry.
 For that purpose we have looked 
for particular channels whose branching ratios were enhanced in supersymmetry
 with respect to their predictions in the SM and in the 2HDMII. If such processes are finally seen in 
 any of the next generation colliders, they could provide interesting clues to
 physics beyond the SM and the 2HDMII.

In order to discern between the MSSM and the  non-SUSY 2HDMII,
 one has to search beyond tree level, where the 
contributions of SUSY particles, via radiative corrections, change the
 predictions of the observables. In particular, in this work we have focused 
 on those possible differences at one loop level in the neutral MSSM Higgs
  boson FC decays into second and third generation quarks, namely: 
  $H_a \to b \bar s , s \bar b$ with $H_a = h_o, H_o, A_o$  and 
  $H_a \to t \bar c , c \bar t$ with $H_a = H_o, A_o$. These are induced 
  mainly by loops of squarks and gluinos, as they are the dominant ones since 
  their size is governed by $\alpha_S$.

In order to study these loop induced flavour changing effects, we assumed here 
the most general hypothesis of misalignment, where no extra symmetry has been 
included to simultaneously diagonalize the quark and squark sectors, leading to non-diagonal
squark mass matrices. In this work
we only keep the well motivated intergenerational mixing between 
$\tilde c_L$ and $\tilde t_L$ for the $up$-type squarks, and $\tilde s_L$ and 
$\tilde b_L$ for the $down$-type squarks, parametrized 'a la Sher' by an unique parameter 
$\lambda$ which we take to be $0 \leq \lambda \leq 1$. We performed the complete 
analytical calculation of the FC partial widths and studied numerically the 
size of these loop induced FCHD as a function of our simplified choice for 
the MSSM parameters, namely $m_A$, $\tan \beta$, $\mu$, $M_{\tilde g}$, 
$M_o$ and $A$. In order to understand the behaviour of $\Gamma (H_a \to b \bar s + s \bar b)$ and  
$\Gamma (H_a \to t \bar c + c \bar t)$ in 
different regions of the MSSM parameter space, we have analyzed in full detail 
the dependence of such FC processes with all 
these MSSM parameters, being scanned by pairs, and for a fixed $\lambda$ value. 
After that, we have studied as well the behaviour with $\lambda$, by varying 
it in the range $0 \leq \lambda \leq 1$.

From this numerical analysis we have learned that the $bs$ decays grow with $\mu$ and $\tan \beta$,
being almost independent on the value of the parameter $A$, while the $tc$ decays have a very
mild dependence on $\tan \beta$, decreasing slowly with it, and the parameter $A$ acquires 
more importance. Concerning the behaviour of the decay widths with $M_o$ we found that all of them
decrease as $M_o$ grows. Notice that we have been very careful when choosing different values for 
$\mu$, $\tan \beta$, $A$ and $M_o$ since some choices do lead to too low values of the squark masses,
being below our required value of $150 \, GeV$.

Looking at the behaviour with $m_A$, we saw that due to the obvious phase 
space effect, the FC decay widths for $H_o$ and $A_o$ clearly grow with 
$m_A$. In contrast, since the lightest Higgs mass has an upper limit, the FC decays involving 
$h_o$ start growing with $m_A$ but then decrease with it for large $m_A$ values, 
manifesting the setting 
of the expected decoupling behaviour. On the other hand, the 
phenomenologically interesting value of $m_A$ would be one that can allow the 
next generation colliders to detect and study all the three neutral Higgs 
bosons.

Finally, the behaviour of all the decay widths under study with $M_{\tilde g}$ 
is clear, 
they first grow with this parameter, reach a maximum, and then they suffer a 
slow decoupling.

With this complete analysis, we ended up by selecting some points of the 
MSSM parameter space that belong to the region where the these FC effects are 
maximized. Then we studied the behaviour of the 
corresponding branching ratios with the parameter $\lambda$ in the previously 
selected region, obtaining  quite sizable rates, whose implications at next 
generation colliders would be interesting to further study in the future. 
Concretely, for $|\lambda| \lsim 0.6$, and for the selected points in the 
MSSM parameter space specified in section 4, 
we have found the following approximate 
ratios $Br (h_o \to b \bar s + s \bar b) \lsim 0.01$, 
$Br (H_o, A_o \to b \bar s + s \bar b) \lsim 0.2$ and 
$Br(H_o, A_o \to t \bar c + c \bar t) \lsim 0.00005$. 

Finally, we studied the behaviour of these FC decay processes in the most 
pessimistic scenario where a very heavy supersymmetric spectrum was 
considered and we showed that the size of these FC effects remain sizable 
even for squark and gluino masses as large as $\mathcal{O}(1TeV)$. 
We have shown in full detail by an explicit analytical computation that 
this unexpected large size of the FC effects is due to the non-decoupling 
behaviour of the heavy squarks and gluinos in the loops. In fact, we have 
presented the asymptotical results corresponding to large SUSY masses, $M_{S} >> M_{EW}$, for 
the FC form factors (and hence, the corresponding FC effective couplings, 
$H_a b \bar s$ and $H_a t \bar c$), 
which can be of much interest for future estimates of rare processes 
production rates at the next generation colliders. Finally, we have studied 
the behaviour of $\Gamma (h_o \to b \bar s + s \bar b)$
 in the asymptotic limit of very large 
pseudoscalar mass, $m_A >> m_Z$, and we have recovered, as expected,
 the vanishing SM tree-level result. 
 
In conclusion, the one loop induced flavour changing Higgs decays offer 
a promising scenario where to look for indirect signals of supersymmetry. 
These decays get quite sizable contributions in some regions of the MSSM parameter 
space, which remain, due to the non-decoupling behaviour of the squarks and 
gluinos, of considerable size even in the most pessimistic case of a very 
heavy SUSY spectrum. With such analysis in mind, and due to their interesting 
phenomenological implications, it would be crucial to further study the 
experimental possibilities of finding these effects in the next planned 
colliders.

%%%%%%%%%%%%%%%%%%%%%%%%%%%%%%%%%%%%%%%%%%%%%%%%%%%%%%%%%%%%%%%%%%%%%%%%%%%%

\section*{Acknowledgments}
This work has been supported in part by
the Spanish Ministerio de Ciencia y Tecnolog{\'\i}a under projects CICYT
FPA 2000-0980 and FPA2000-3172-E. A.~M.~C. acknowledges MECD for financial support by 
FPU grant AP2001-0521.

%%%%%%%%%%%%%%%%%%%%%%%%%%%%%%%%%%%%%%%%%%%%%%%%%%%%%%%%%%%%%%%%%%%%%%%%%%%%
\begin{center}
{\Large \bf Appendix A}
\end{center}
\appendix
\renewcommand{\theequation}{A.\arabic{equation}}

In this appendix we present the explicit values of the Higgs-squark-squark
couplings in the squark mass eigenstate basis that appear in
eqs.(\ref{formfactorLbs}) and (\ref{formfactorRbs}). 
For the  $up$-type squarks, these couplings are as follows:
\begin{eqnarray}
g_{H_a \tilde u_{\alpha} \tilde u_{\beta}}& = & i \left[ \frac{g M_Z}{\cos 
\theta_W} \left( V_{1,a}^u \left[ ( \frac{1}{2} - \frac{2}{3} \sin^2 \theta_W)
[R_{1 \alpha}^{u*} R_{1 \beta}^u + R_{3 \alpha}^{u*} R_{3 \beta}^u]
\right. \right. \right. \nn \\
&& \left. \left. + \frac{2}{3} \sin ^2 \theta_W [R_{2 \alpha}^{u*} R_{2 \beta}^u + 
R_{4 \alpha}^{u*} R_{4 \beta}^u] \right] \right) - \frac{g m_c^2}{M_W \sin \beta} 
V_{2,a}^u [R_{1 \alpha}^{u*} R_{1 \beta}^u + R_{2 \alpha}^{u*} R_{2 \beta}^u]
\nn \\
&& - \frac{g m_t^2}{M_W \sin \beta} V_{2,a}^u [R_{3 \alpha}^{u*} R_{3 \beta}^u + 
R_{4 \alpha}^{u*} R_{4 \beta}^u] - \frac{g m_c}{ 2 M_W \sin \beta} V_{3,a}^u 
[R_{1 \alpha}^{u*} R_{2 \beta}^u] 
\nn \\
&&   - \frac{g m_t}{ 2 M_W \sin \beta} V_{3,a}^u [R_{3 \alpha}^{u*} R_{4 \beta}^u]  
- \frac{g m_c}{ 2 M_W \sin \beta} V_{4,a}^u [R_{2 \alpha}^{u*} R_{1 \beta}^u] 
\nn \\
&& \left. - \frac{g m_t}{ 2 M_W \sin \beta} V_{4,a}^u [R_{4 \alpha}^{u*} 
R_{3 \beta}^u] \right]
\label{ghsusu1}
\end{eqnarray}
with:
\begin{eqnarray}
V_{1,a}^u &=& \left( \sin (\alpha + \beta), - \cos (\alpha + \beta), 0 \right),\nn \\
V_{2,a}^u &=& \left( \cos (\alpha), \sin (\alpha), 0 \right), 
\nn \\
V_{3,a}^u &=& \left( A_q \cos (\alpha) + \mu \sin (\alpha), A_q \sin (\alpha) - \mu \cos (\alpha), (A_q \cos \beta + \mu \sin \beta)/i \right), 
\nn \\
V_{4,a}^u &=& \left( A_q \cos (\alpha) + \mu \sin (\alpha), A_q \sin (\alpha) - \mu \cos (\alpha), - (A_q \cos \beta + \mu \sin \beta)/i \right)\nn \\  
\label{ghsusu2}
\end{eqnarray} 
for $H_a=(h_o,H_o,A_o)$, respectively, and where $A_q = A_t, A_c$
correspondingly.
 
Similarly, for the $down$-type squarks, the couplings are as follows:
\begin{eqnarray}
g_{H_a \tilde d_{\alpha} \tilde d_{\beta}}& = & i \left[ \frac{g M_Z}{\cos 
\theta_W} \left( V_{1,a}^d \left[ ( \frac{1}{2} - \frac{1}{3} \sin^2 \theta_W)
[R_{1 \alpha}^{d*} R_{1 \beta}^d + R_{3 \alpha}^{d*} R_{3 \beta}^d]
\right. \right. \right. \nn \\
&& \left. \left. + \frac{1}{3} \sin ^2 \theta_W [R_{2 \alpha}^{d*} R_{2 \beta}^d + 
R_{4 \alpha}^{d*} R_{4 \beta}^d] \right] \right) - \frac{g m_s^2}{M_W \cos \beta} 
V_{2,a}^d [R_{1 \alpha}^{d*} R_{1 \beta}^d + R_{2 \alpha}^{d*} R_{2 \beta}^d]
\nn \\
&& - \frac{g m_b^2}{M_W \cos \beta} V_{2,a}^d [R_{3 \alpha}^{d*} R_{3 \beta}^d + 
R_{4 \alpha}^{d*} R_{4 \beta}^d] - \frac{g m_s}{ 2 M_W \cos \beta} V_{3,a}^d 
[R_{1 \alpha}^{d*} R_{2 \beta}^d] 
\nn \\
&&   - \frac{g m_b}{ 2 M_W \cos \beta} V_{3,a}^d [R_{3 \alpha}^{d*} R_{4 \beta}^d]  
- \frac{g m_s}{ 2 M_W \cos \beta} V_{4,a}^d [R_{2 \alpha}^{d*} R_{1 \beta}^d] 
\nn \\
&& \left. - \frac{g m_b}{ 2 M_W \cos \beta} V_{4,a}^d [R_{4 \alpha}^{d*} 
R_{3 \beta}^d] \right]
\label{ghsdsd1}
\end{eqnarray}
with:
\begin{eqnarray}
V_{1,a}^d &=& \left(- \sin (\alpha + \beta),  \cos (\alpha + \beta), 0 \right),\nn \\
V_{2,a}^d &=& \left( -\sin (\alpha), \cos (\alpha), 0 \right), 
\nn \\
V_{3,a}^d &=& \left( - A_q \sin (\alpha) - \mu \cos (\alpha), A_q \cos (\alpha) - \mu \sin (\alpha), (A_q \sin \beta + \mu \cos \beta)/i \right), 
\nn \\
V_{4,a}^d &=& \left( - A_q \sin (\alpha) - \mu \cos (\alpha), A_q \cos (\alpha) - \mu \sin (\alpha), - (A_q \sin \beta + \mu \cos \beta)/i \right)\nn \\  
\label{ghsdsd2}
\end{eqnarray}
for $H_a=(h_o,H_o,A_o)$, respectively, and 
where $A_q = A_b, A_s$ correspondingly.

In the previous formulas,  
$R^u_{\alpha \beta}$ and $R^d_{\alpha \beta}$ with $\alpha, \beta=1,2,3,4$, 
are the rotation matrices that diagonalize the squark 
squared mass matrices of eq.(\ref{eq.usquarkmass}) and (\ref{eq.dsquarkmass}) 
respectively.
%%%%%%%%%%%%%%%%%%%%%%%%%%%%%%%%%%%%%%%%%%%%%%%%%%%%%%%%%%%%%%%%%%
 \begin{center}
{\Large \bf Appendix B}
\end{center}
\appendix
\setcounter{equation}{0}
\renewcommand{\theequation}{B.\arabic{equation}}

In this appendix we give the expressions required to compute the leading contribution to the 
FCHD partial widths in the large SUSY mass expansion defined in Sect. 5.
For that purpose we first write the values of the squark masses and rotation matrices and then
the formulae for the two- and three-point integrals.

The expressions for the squark masses and rotation matrices, in the limit of large SUSY mass
parameters and keeping just the leading contribution, are

\begin{eqnarray}
&M_{\tilde q_1}^2 \simeq M_o^2 (1+\lambda) 
\, \, , \, \,
M_{\tilde q_2}^2 \simeq M_o^2 \, , \,
M_{\tilde q_3}^2 \simeq M_o^2  \, \, , \, \,
M_{\tilde q_4}^2 \simeq M_o^2 (1-\lambda) &
\end{eqnarray}
\begin{eqnarray}
&R_{13}^{(d)} \simeq - R_{12}^{(d)} \simeq R_{44}^{(d)} \simeq R_{41}^{(d)} 
\frac{m_b}{\sqrt{2}\lambda M_o^2} (A- \mu \tan\beta) &\nn \\
&R_{14}^{(d)} \simeq R_{11}^{(d)} \simeq - R_{23}^{(d)} \simeq - R_{22}^{(d)}
- R_{34}^{(d)} \simeq R_{31}^{(d)} \simeq - R_{43}^{(d)} \simeq R_{42}^{(d)}
\simeq \frac{1}{\sqrt{2}} & \nn \\
&R_{24}^{(d)} \simeq - R_{21}^{(d)} \simeq - R_{33}^{(d)} \simeq - R_{32}^{(d)}
\simeq - \frac{m_s}{\sqrt{2}\lambda M_o^2} (A-\mu \tan\beta) &   
\end{eqnarray}
with similar results for $R^{(u)}$ just replacing $b \to t$, $s \to c$ and 
$\tan\beta \to \cot\beta$.

In this limit the expressions for the two- and three-point 
one-loop integrals involved 
are,
\bea
\label{eq.intleading}
&& C_0(m_q^2,m_H^2,m_{q'}^2;M_{\tilde g}^2, M_{\tilde q_a}^2, 
M_{\tilde q_b}^2) \simeq -\frac{1}{2 M_o^2} f_1 (R_{q_a}, R_{q_b}) + \mathcal{O}\left(
\frac{M_{EW}}{M_o^3} \right) \nn\\
&& C_{11}(m_q^2,m_H^2,m_{q'}^2;M_{\tilde g}^2, M_{\tilde q_a}^2,
    M_{\tilde q_b}^2) \simeq \frac{1}{3 M_o^2} f_{10} (R_{q_a},R_{q_b}) + \mathcal{O}\left(
\frac{M_{EW}}{M_o^3} \right) \nn \\
&& C_{12}(m_q^2,m_H^2,m_{q'}^2;M_{\tilde g}^2, M_{\tilde q_a}^2,
M_{\tilde q_b}^2) \simeq \frac{1}{6 M_o^2} f_{13} (R_{q_a},R_{q_b})+ \mathcal{O}\left(
\frac{M_{EW}}{M_o^3} \right)  \nn\\
&&B_0(m_q^2;M_{\tilde q_a}^2,M_{\tilde g}^2)
    \simeq  \Delta - log\frac{M_{\tilde q_a}^2}{\mu_0^2} + g_1 (R_{q_a})
+ \mathcal{O}\left(
\frac{M_{EW}}{M_o} \right) \nn \\ 
&&B_1(m_q^2;M_{\tilde q_a}^2,M_{\tilde g}^2)
    \simeq - \frac{1}{2} \Delta + \frac{1}{2} log\frac{M_{\tilde q_a}^2}{\mu_0^2}  
+ g_2 (R_{q_a}) + \mathcal{O}\left(
\frac{M_{EW}}{M_o} \right), \nn \\ 
\label{eq.expintegrals}
\eea
where $R_{q_a} = M_{\tilde g}/M_{\tilde q_a}$ and the explicit
 expressions for the functions 
$f_i$ and $g_i$ can be found in~\cite{HaberTemes,ourHtb}.
For the simplest case, where $R_{q_a}=R_{q_b}=1$, they are 
$f_1(1,1)=f_{10}(1,1)=f_{13}(1,1)=1$ and $g_1(1)=g_2(1)=0$.

%%%%%%%%%%%%%%%%%%%%%%%%%%%%%%%%%%%%%%%%%%%%%%%%%%%%%%%%%%%%%%%%%%%%%%%%%%%%

\end{document}